\documentclass[10pt,aps,prd,floats,floatfix,preprintnumbers,superscriptaddress,nofootinbib]{revtex4-2}

\usepackage{graphicx}
\usepackage{epstopdf} 
\usepackage{dcolumn}
\usepackage{amssymb}
\usepackage{amsmath}
\usepackage{bbold}
\usepackage{ulem}
\usepackage[hidelinks]{hyperref}
\usepackage{color}
\usepackage{xcolor,colortbl}
\usepackage[utf8]{inputenc}
\usepackage{slashed}

\newcommand{\be}{\begin{equation}}
\newcommand{\ee}{\end{equation}}

\newcommand{\rmd}{\mathrm{d}}
\newcommand{\rme}{\mathrm{e}}
\newcommand{\rmi}{\mathrm{i}}

\newcommand{\calP}{\mathcal{P}}

\newcommand{\calS}{\mathcal{S}}
\newcommand{\calO}{\mathcal{O}}

\newcommand{\qonp}{\boldsymbol{q}_{1}}
\newcommand{\qtwp}{\boldsymbol{q}_{2}}
\newcommand{\wpe}{\boldsymbol{w}}
\newcommand{\aperp}{\boldsymbol{a}}
\newcommand{\bp}{\boldsymbol{b}}
\newcommand{\cperp}{\boldsymbol{c}}
\newcommand{\xp}{\boldsymbol{x}}
\newcommand{\yp}{\boldsymbol{y}}
\newcommand{\zp}{\boldsymbol{z}}
\newcommand{\qp}{\boldsymbol{q}}
\newcommand{\kp}{\boldsymbol{k}}

\newcommand{\delp}{\boldsymbol{\Delta}}

\newcommand{\kgp}{\boldsymbol{k}_{g}}

\begin{document}
%\date{\today}
\preprint{ZTF-EP-26-04}

\title{Generalized transverse momentum distributions at small-\ensuremath{x}}

\author{Sanjin Beni\' c}
\affiliation{Department of Physics, Faculty of Science, University of Zagreb, Bijenička c. 32, 10000 Zagreb, Croatia}
\author{Yoshikazu Hagiwara}
\affiliation{Department of Physics, Faculty of Science, University of Zagreb, Bijenička c. 32, 10000 Zagreb, Croatia}
\author{Boris \v Sari\' c}
\affiliation{Department of Physics, Faculty of Science, University of Zagreb, Bijenička c. 32, 10000 Zagreb, Croatia}
\author{Eric Andreas Vivoda}
\affiliation{Department of Physics, Faculty of Science, University of Zagreb, Bijenička c. 32, 10000 Zagreb, Croatia}

\begin{abstract}
We compute the complete set of the leading-twist gluon and sea-quark generalized transverse momentum distributions (GTMDs) in the small-$x$, or eikonal, approximation at vanishing skewness $\xi = 0$. All the gluon GTMDs become expressed in terms of the basic gluon dipole operator featuring also proton helicity-flips. Consequently, we establish universal relations between otherwise distinct GTMDs that hold in the small-$x$ limit. 
%{\color{orange}All the otherwise independent gluon GTMDs become expressed in terms of the basic gluon dipole operator, consequently providing universal relations between them in the small-$x$ limit.} 
The obtained results are systematically projected onto the transverse momentum dependent distributions (TMD) and the generalized parton distribution (GPD) cases, recovering known results where available. In case of sea-quarks all the GTMDs are given in terms of the gluon dipole convoluted with a hard kernel. We generalize the unpolarized sea-quark GTMD to non-zero momentum transfers and find new results for the proton helicity-flip distributions. We pay special attention to their perturbative, high transverse momentum, tails to confirm that it becomes governed by the small-$x$ gluon GPDs. The obtained results provide guidance to the phenomenological modeling of the GTMDs but also allow for their explicit computations.
\end{abstract}

\maketitle

\section{Introduction}

Generalized transverse momentum distributions (GTMDs) \cite{Ji:2003ak,Belitsky:2003nz} are non-perturbative objects that provide the most general partonic description of the hadron structure. They encode information on the intrinsic longitudinal and transverse momenta of the underlying partons, together with the longitudinal and transverse momentum transfer between the initial and final hadron states, thereby enabling spatial imaging of partons. The complete leading-twist classification of the quark and gluon GTMDs has been developed in Refs.~\cite{Meissner:2009ww,Lorce:2013pza}, hosting a total of 16 independent quark and 16 independent gluon GTMDs.
Following \cite{Lorce:2013pza}, the GTMDs can be classified according to their multipole patterns, that is the orbital angular momentum transfers from the proton to the active parton system. For gluon GTMDs these are $S$, $P$, $D$ and $F$-wave GTMDs at the leading twist. The quark GTMDs are classified into $S$, $P$ and $D$-waves at the leading twist.
%{\color{orange}Following \cite{Lorce:2013pza}, they are characterized according to their orbital angular momentum transfer from the proton. In the gluon case these are the $S$, $P$, $D$, and $F$-wave GTMDs, while in the sea-quark case these are $S$, $P$, and $D$-waves.}

Experimental access to the GTMDs may be possible at the future Electron-Ion Colliders (EICs) \cite{Accardi:2012qut,AbdulKhalek:2021gbh} or in the ultra-peripheral collisions \cite{Baltz:2007kq,Klein:2017vua}, and potential signatures have been put forward in many works, see e.~g.~\cite{Hatta:2016dxp,Hagiwara:2017fye,Hatta:2016aoc,Bhattacharya:2017bvs,Bhattacharya:2018lgm,Boer:2021upt,Bhattacharya:2023yvo,Bhattacharya:2023hbq,Benic:2023ybl,Hagiwara:2020juc,Bhattacharya:2026qnd}. However, it seems fair to say that the sheer number of independent and complex-valued functions involved, as well as their multi-dimensional kinematic dependencies and non-trivial relationships to the actual observables, makes their experimental extractions challenging. In this light, we want to draw attention to the high-energy, or small-$x$, limit in which case the gluon GTMDs are all controlled by the so called the gluon dipole distribution \cite{Dominguez:2011wm}. Being a scalar function, the gluon dipole holds fewer independent structures than the original GTMD decomposition itself. By such a simple argument the hope is that this particular kinematic limit will at least bring down the number of independent GTMDs clarifying the phenomenology for collisions at high-energies and guide their extractions beyond it. This was recognized early on in the transverse momenutum distribution (TMD) limit establishing an equality between the unpolarized and the linearly polarized gluon TMDs at small-$x$ \cite{Metz:2011wb,Dominguez:2011br}. In \cite{Boer:2015pni,Boer:2016xqr}, a similar connection was found between the gluon TMDs of the transversely polarized proton. Later the generalization to the GTMDs was made \cite{Boer:2018vdi}, but only in case of unpolarized proton, see also \cite{Bhattacharya:2025fnz}. Polarization effects were studied more recently in Ref.~\cite{Bhattacharya:2024sno} that revealed an interesting connection between the spin-orbit and the unpolarized GTMD. Notwithstanding these developments, a complete and systematic analysis of all possible connections between the small-$x$ gluon GTMDs covering the off-forward ($\delp \neq 0$) kinematics and proton helicity-flips has not been made. This is one of the motivations for this work. Another motivation is to express all the GTMDs by the common small-$x$ objects: the Pomeron \cite{Balitsky:1978ic,Mueller:1993rr} and the Odderon \cite{Lukaszuk:1973nt} through their realizations as real and imaginary parts of the gluon dipole distributions \cite{Kovchegov:2012mbw}, thereby making a connection between the GTMD and small-$x$ formalisms.

In Sec.~\ref{sec:ggtmd} we show that $S$-wave and $D$-wave gluon GTMDs are both associated with helicity non-flip (or spin-independent) gluon dipoles, while $P$ and $F$-wave gluon GTMDs are associated with helicity flip (or spin-dependent) dipoles. This turns into relations between the $S$-wave and the $D$-wave gluon GTMDs and between the $P$- and the $F$-wave gluon GTMDs. We also find relations for gluon GTMDs that differ by the gluon helicity transfer. In the TMD limit all the known results from the literatures \cite{Metz:2011wb,Boer:2015pni,Boer:2016xqr} are recovered. Finally, the generalized parton distribution (GPD) limit is analyzed, connecting all the gluon GPDs to the real part of the dipole distribution (the Pomerons) clarifying the roles of the isotropic and the elliptic parts. The imaginary parts of the dipole (the Odderons) are connected to a class of twist-3 tri-gluon GPDs.

In the same vein, the sea-quark unpolarized TMD \cite{Marquet:2009ca,Xiao:2017yya} and the sea-quark Sivers function \cite{Dong:2018wsp,Kovchegov:2021iyc} have both been computed at small-$x$ where they become expressed in terms of the real and the imaginary part of the gluon dipole distributions, respectively. In the case of sea-quark GTMDs, Ref.~\cite{Bhattacharya:2024sno} found a relation between the unpolarized and the spin-orbit GTMD and very recent works are addressing skewness dependence for the unpolarized GTMD \cite{Bhattacharya:2025fnz,Kovchegov:2025yyl}. In Sec.~\ref{sec:seaq} we compute the complete set of sea-quark GTMDs at small-$x$. Similar to the gluon case we find the $S$-wave ($P$-wave) quark GTMDs to be related to the spin-independent (spin-dependent) dipole distributions. The obtained results are studied in the high-$\kp$ limit, where we were able to factorize all the obtained general results in terms of the gluon GPDs covering both the real and the imaginary parts of the GTMDs as well as the $\delp \neq 0$ corrections. In the final Sec.~\ref{sec:concl} we summarize our results. Some of the computations steps from the main Secs.~\ref{sec:ggtmd} and \ref{sec:seaq} are provided in the appendices. 

\section{Gluon GTMDs}
\label{sec:ggtmd}

Using the conventions from \cite{Lorce:2013pza}, our starting point is the gluon GTMD correlator
\be
W^{ij}_{\Lambda' \Lambda}(x,\xi,\kp,\delp) = \frac{2}{x P^+} \int \frac{\rmd z^- \rmd^2 \zp}{(2\pi)^3} \rme^{\rmi x P^+ z^- - \rmi \kp \cdot \zp} \langle p' \Lambda' |{\rm tr}\left(F^{+i}(-z/2) \mathcal{W} F^{+j}(z/2)\mathcal{W}'\right) | p \Lambda\rangle\,.
\label{eq:ggtmd}
\ee
Here $\delp = \boldsymbol{p}' - \boldsymbol{p}$, $P^+ = (p^+ + p'^+)/2$, $\xi = -(p'^+ - p^+)/(2P^+)$. We will be interested in the so-called dipole-type configuration of the gauge link where $\mathcal{W}$ ($\mathcal{W}'$) extends from $-z/2$ ($+z/2$) to $+\infty$ ($-\infty$) and then to $+z/ 2$ ($-z/2$). An alternative possibility is the Weisz\" acker-Williams type distribution \cite{Bomhof:2007xt,Dominguez:2011wm}, and the choice between the two depends on the physical process. 

%{\color{blue} In this work, we adopt the conventions of \cite{Lorce:2013pza}. Our starting point is the gluon GTMD correlator, defined as 
%\be
%W^{ij}_{\Lambda' \Lambda}(x,\xi,\kp,\delp) = \frac{2}{x P^+} \int \frac{\rmd z^- \rmd^2 \zp}{(2\pi)^3} \rme^{\rmi x P^+ z^- - \rmi \kp \cdot \zp} \langle p' \Lambda' |{\rm tr}\left(F^{+i}(-z/2) \mathcal{W} F^{+j}(z/2)\mathcal{W}'\right) | p \Lambda\rangle\,.
%\label{eq:ggtmd}
%\ee
%Here $\delp = \boldsymbol{p}' - \boldsymbol{p}$, $P^+ = (p^+ + p'^+)/2$, $\xi = (p^+ - p'^+)/(2P^+)$, %, {\color{orange} $x=k^+/P^+$}
%while $\mathcal{W}$ and $\mathcal{W}'$ denote gauge links connecting the points $-z/2$ and $z/2$ through the light-cone infinity $z^- = \pm \infty$. We focus on the so-called dipole-type configuration of the gauge link and the underlying dipole-type gluon distribution. An alternative option is the  Weisz\" acker-Williams type distribution \cite{}, and the appropriate choice between the two depends on the physical process under consideration.}

The different components of $W^{ij}$ are gathered into a helicity matrix in the joint proton ($\Lambda = \pm 1/2$) and gluon ($\lambda = \pm 1$) helicity space
\be
H^g_{\Lambda' \lambda',\Lambda \lambda} = 
\begin{pmatrix}
\frac{1}{2}(U^g + L^g)_{\Lambda'\Lambda} & \frac{1}{2}T^g_{R,\Lambda'\Lambda}\\
 \frac{1}{2}T^g_{L,\Lambda'\Lambda} & \frac{1}{2}(U^g - L^g)_{\Lambda'\Lambda}\\
\end{pmatrix}_{\lambda'\lambda}\,.
\label{eq:Hg}
\ee
with the individual contributions labelled as $U^g \equiv \delta^{ij} W^{ij}$, $L^g \equiv -\rmi \epsilon^{ij} W^{ij}$, $T^g_R \equiv - R^i R^j W^{ij}$, $T^g_L \equiv - L^i L^j W^{ij}$, and $\boldsymbol{L} = (1,-\rmi)$ and  $\boldsymbol{R} = (1,\rmi)$ are the helicity vectors. The rows (columns) of \eqref{eq:Hg} are labeled with $\lambda$ ($\lambda'$).
The individual matrix element of $H^g_{\Lambda'\lambda',\Lambda\lambda}$ have the decomposition into a total of 16 GTMDs at the leading twist \cite{Lorce:2013pza}, spanning the possibility of no helicity flip or helicity flip on the proton side only, gluon side only or both. Orbital angular momentum transfer is fixed as $\Delta l_z = \Lambda - \lambda - (\Lambda' - \lambda')$. The GTMDs are collectively labeled through its $\Delta l_z = S$, $P$, $D$, $F$, $\dots$, spin transfer $\Delta S_z$, light-front (LF) parity $c_P$ and twist $t+1$ as $(\Delta l_z)_{t}^{\Delta S_z,c_P}(x,\xi,\kp,\delp)$. The transverse momentum dependence is written as $(\kp, \delp)$, while more precisely, this implies a set of three independent variables: $(\kp^2,\delp^2,\kp\cdot\delp)$.  In the case when there is no helicity flip on either the proton or the gluon side,  we have
\be
\begin{split}
& H^g_{+\frac{1}{2} +1, +\frac{1}{2} + 1} = \frac{1}{2}\left[(S^{0,+;g}_{1,1a} + S^{0,-;g}_{1,1a}) + \rmi \frac{\kp\times\delp}{M^2}(S^{0,+;g}_{1,1b} + S^{0,-;g}_{1,1b})\right]\,,\\
& H^g_{-\frac{1}{2} +1, -\frac{1}{2} + 1} = \frac{1}{2}\left[(S^{0,+;g}_{1,1a} - S^{0,-;g}_{1,1a}) - \rmi \frac{\kp\times\delp}{M^2}(S^{0,+;g}_{1,1b} - S^{0,-;g}_{1,1b})\right]\,.
\end{split}
\label{eq:GTMD1}
\ee
For brevity, we will often be dropping the kinematical dependence $(x,\xi,\kp,\delp)$ of the GTMDs as in \eqref{eq:GTMD1}. Here $\aperp\times\bp = \epsilon^{ij}a^i b^j = a^1 b^2 - a^2 b^1$. The two other matrix elements, $H^g_{-\frac{1}{2}-1,-\frac{1}{2}-1}$ and $H^g_{+\frac{1}{2}-1,+\frac{1}{2}-1}$, are obtained by LF parity which flips the helicity and the $a^1$ component of some transverse vector $\aperp$. Since there is no proton nor parton helicity transfer in \eqref{eq:GTMD1}, it is characterized by $\Delta l_z = 0$, determining the two possible structures: $1$ and $\kp\times\delp$, and with the corresponding GTMDs labeled as $S$-wave ($S_a$ and $S_b$). Within these two structures we have two different scalar functions $S^{0,+;g}_{1,1a}$, $S^{0,-;g}_{1,1a}$ (and same for $S_b$) allowing for the possibility that the parton can carry helicity opposite in sign to that of the proton that is captured by parity-odd distributions. We can isolate a particular GTMD in \eqref{eq:GTMD1} by a linear combination of the helicity matrix. For example,
\be
\begin{split}
&\frac{1}{2}\sum_{\lambda = \pm 1} H^g_{+\frac{1}{2}\lambda,+\frac{1}{2}\lambda} + \frac{1}{2}\sum_{\lambda = \pm 1} H^g_{-\frac{1}{2}\lambda,-\frac{1}{2}\lambda} = S_{1,1a}^{0,+;g}\,,\\
&\frac{1}{2}\sum_{\lambda = \pm 1} H^g_{+\frac{1}{2}\lambda,+\frac{1}{2}\lambda} - \frac{1}{2}\sum_{\lambda = \pm 1} H^g_{-\frac{1}{2}\lambda,-\frac{1}{2}\lambda} = \rmi (\kp\times\delp) S_{1,1b}^{0,+;g}\,,
\end{split}
\ee
clarifies $S_{1,1a}^{0,+;g}$ as the unpolarized GTMD and $S^{0,+;g}_{1,1b}$ as the GTMD associated with orbital angular momentum that appears in the polarized proton \cite{Lorce:2011kd}. Similarly, $S_{1,1a}^{0,-;g}$ is understood as the GTMD related to the helicity distribution and $S_{1,1b}^{0,-;g}$ as the GTMD related to the spin-orbit distribution \cite{Lorce:2011kd,Bhattacharya:2024sck}.

Turning to the case where helicity transfer occurs on the proton side only, we have
\be
\begin{split}
& H^g_{+\frac{1}{2} +1, -\frac{1}{2} + 1} = \frac{1}{2}\left[-\frac{k_L}{M}(P^{0,+;g}_{1,1a} - P^{0,-;g}_{1,1a}) - \frac{\Delta_L}{M}(P^{0,+;g}_{1,1b} - P^{0,-;g}_{1,1b})\right]\,,\\
& H^g_{-\frac{1}{2} +1, +\frac{1}{2} + 1} = \frac{1}{2}\left[\frac{k_R}{M}(P^{0,+;g}_{1,1a} + P^{0,-;g}_{1,1a}) + \frac{\Delta_R}{M}(P^{0,+;g}_{1,1b} + P^{0,-;g}_{1,1b})\right]\,,
\end{split}
\label{eq:GTMD2}
\ee
that is accompanied by a single unit of angular momentum transfer through the appearance of the vectors\footnote{In general, the angular structure $k_R^N k_L^n \Delta_R^M \Delta_L^m$ carries the angular momentum quantum number $N-n+M-m$.} $k_{R,L} = k^1 \pm \rmi k^2$ and $\Delta_{R,L} = \Delta^1 \pm \rmi \Delta^2$. Summing over the gluon helicities selects a linear combination $\sim k_{R,L} P_{1,1a}^{0,+;g} + \Delta_{R,L} P_{1,1b}^{0,+;g}$. Passing to the transverse spin basis \cite{Meissner:2007rx} we can interpret it as an off-forward generalization of the gluon Sivers function\footnote{In the forward limit the connection is given in the third line of \eqref{eq:tmdl}.} that describes unpolarized gluons in a transversely polarized proton.

In case of helicity transfer on the gluon side, we need two units of angular momentum transfer and so the resulting decomposition reads
\be
\begin{split}
& H^g_{+\frac{1}{2} +1, +\frac{1}{2} - 1} = -\frac{1}{2}\left[\frac{k_R^2}{M^2}(D^{2,+;g}_{1,1a} + D'^{2,+;g}_{1,1a}) + \frac{\Delta_R^2}{M^2}(D^{2,+;g}_{1,1b} + D'^{2,+;g}_{1,1b})\right]\,,\\
& H^g_{-\frac{1}{2} +1, -\frac{1}{2} - 1} = -\frac{1}{2}\left[\frac{k_R^2}{M^2}(D^{2,+;g}_{1,1a} - D'^{2,+;g}_{1,1a}) + \frac{\Delta_R^2}{M^2}(D^{2,+;g}_{1,1b} - D'^{2,+;g}_{1,1b})\right]\,.
\end{split}
\label{eq:GTMD3}
\ee
One could also consider a mixed structure of the type $k_R \Delta_R$ as it carries same angular momentum. However, $k_R \Delta_R$ is not independent structure and can be expressed as linear combination of $k_R^2$ and $\Delta_R^2$ - see App.~\ref{sec:inner}. It is completely conventional \cite{Lorce:2013pza} to use $k_R^2$ and $\Delta_R^2$ only: a decomposition in terms of $k_L^2$ and $\Delta_L^2$ is connected to \eqref{eq:GTMD3} by a LF parity transformation. Summing over the proton helicities brings a combination $\sim k_R^2 D^{2,+;g}_{1,1a} + \Delta_R^2 D^{2,+;g}_{1,1b}$ that can be understood as an off-forward generalization of the gluonic variant of the Boer-Mulders function \cite{Mulders:2000sh}\footnote{In the forward limit, the connection is given in the fourth line of \eqref{eq:tmdl}.}.

Finally, in case both proton helicity and gluon helicity is flipped we can have one unit or three units of change of orbital angular momentum leading to $P$-wave and $F$-wave distributions
\be
\begin{split}
& H^g_{+\frac{1}{2} +1, -\frac{1}{2} - 1} = -\frac{1}{2}\left[\frac{k_R}{M}P^{2,+;g}_{1,1a} + \frac{\Delta_R}{M}P^{2,+;g}_{1,1b}\right]\,,\\
& H^g_{-\frac{1}{2} +1, +\frac{1}{2} - 1} = -\frac{1}{2}\left[\frac{k_R^3}{M^3}F^{2,+;g}_{1,1a} + \frac{\Delta_R^3}{M^3}F^{2,+;g}_{1,1b}\right]\,,
\end{split}
\label{eq:GTMD4}
\ee
that can be understood as generalizations of gluonic transversity GPDs, describing transversely polarized gluons in transversely polarized protons - the precise connection to the GPDs is given in \eqref{eq:Gpd1}.

\subsection{Matching to the gluon dipole GTMD at small-\ensuremath{x}}

Now we show that in the small-$x$, or eikonal, approximation, the results in \eqref{eq:GTMD1}-\eqref{eq:GTMD4} can be written down in terms of the so-called gluon dipole distribution - a basic object in the small-$x$ physics that is expressed entirely in terms of the longitudinal Wilson lines $V(\xp) = P \exp\left[-\rmi g \int_{-\infty}^\infty \rmd x^+ A^-(x^+,\xp)\right]$. For this purpose we use the result that at small-$x$ (and at $\xi = 0$) $W^{ij}_{\Lambda'\Lambda}$ can be written as \cite{Dominguez:2011wm,Hatta:2016dxp,Boer:2018vdi}
\be
W_{\Lambda'\Lambda}^{ij}(x,\xi = 0,\kp,\delp) = \frac{2 N_c}{x\alpha_S} \left(k^i + \frac{1}{2}\Delta^i\right) \left(k^j - \frac{1}{2}\Delta^j\right) \calS_{\Lambda' \Lambda}(\kp,\delp)\,,
\label{eq:Wsmall}
\ee
where
\be
\calS_{\Lambda'\Lambda}(\kp,\delp) = \int\frac{\rmd^2 \xp}{(2\pi)^2} \int\frac{\rmd^2 \yp}{(2\pi)^2} \rme^{-\rmi \kp\cdot(\xp - \yp) + \rmi \delp\cdot\frac{\xp + \yp}{2}} \frac{1}{N_c} \frac{\langle p'\Lambda' |{\rm tr}\left[ V^\dag(\xp)V(\yp)\right]|p\Lambda\rangle}{\langle P\Lambda|P\Lambda\rangle}\,,
\label{eq:dipgtmd}
\ee
is the dipole GTMD \cite{Hatta:2016dxp,Boer:2018vdi}. Thanks to the unitary property of the Wilson lines, the dipole GTMD satisfies the familiar sum rule
\be
\int \rmd^2 \kp \calS_{\Lambda'\Lambda}(\kp,\delp) = \delta^{(2)}(\delp)\delta_{\Lambda'\Lambda}\,,
\label{eq:sumr}
\ee
which holds separately for its real and imaginary parts.

As the matching \eqref{eq:Wsmall} completely factorizes the gluon helicity indices $ij$, the remaining object $S_{\Lambda'\Lambda}(\kp,\delp)$ depends only on proton helicities and is therefore much simpler, resulting in a parametrization in terms of only three scalar functions \cite{Boussarie:2019vmk,Hagiwara:2020mqb}
\be
\calS_{\Lambda'\Lambda}(\kp,\delp) = \delta_{\Lambda\Lambda'} \calS(\kp,\delp) + \delta_{\Lambda,-\Lambda'} \frac{2\Lambda k^1 + \rmi k^2}{M} \calS^\perp_{1T}(\kp,\delp) + \delta_{\Lambda,-\Lambda'} \frac{2\Lambda \Delta^1 + \rmi \Delta^2}{M} \calS_T(\kp,\delp)\,.
\label{eq:Sdecompose1}
\ee
In the transverse spin ($\boldsymbol{S}$) basis, the helicity flip terms would correspond to $\kp\times \boldsymbol{S}$ and $\delp\times \boldsymbol{S}$-type modulations. Strictly speaking, the scalar functions $\calS$, $\calS_{1T}^\perp$ and $\calS_T$ are constant in $x$ - the $x$ dependence can be introduced through the small-$x$ evolutions \cite{Balitsky:1995ub,Kovchegov:1999yj}. The real and imaginary parts of these complex functions define the Pomerons and the Odderons, respectively, and are decomposed in the following way \cite{Boussarie:2019vmk,Hagiwara:2020mqb}
\be
\begin{split}
& \calS = \calP + \rmi \frac{(\kp\cdot \delp)}{M^2}\calO\,,\\
& \calS_{1T}^\perp = \frac{(\kp\cdot \delp)}{M^2}\calP_{1T}^\perp + \rmi \calO_{1T}^\perp\,,\\
& \calS_T = \calP_T + \rmi \frac{(\kp\cdot \delp)}{M^2}\calO_T\,.
\end{split}
\label{eq:Sdecompose2}
\ee
The above decomposition can be understood on the basis of $C$-parity as Pomerons are $C$-even gluon exchanges, and Odderons are $C$-odd gluon exchanges. For example, ${\rm Re}(\calS)$ (${\rm Im}(\calS)$) is even (odd) in $\kp\to - \kp$ because Pomeron (Odderon) is even (odd) under $\xp \leftrightarrow \yp$, see e.~g.~\cite{Hatta:2005as}. At high energies $C$-parity and LF time reversal transformations \cite{Lorce:2013pza}, that effectively switches $\delp \to -\delp$ in the dipole distribution, are the same \cite{Sievert:2014psa}, and so it is another way to understand the decomposition in \eqref{eq:Sdecompose2}. 

The functions $\calP(\kp,\delp)$ and $\calO(\kp,\delp)$ appearing in \eqref{eq:Sdecompose2} are the helicity-non-flip (or spin-independent) Pomeron and Odderon that are ubiquitous in the small-$x$ computations and appearing in many processes such as exclusive dijet \cite{Hagler:2002nh,Hatta:2016dxp,Boer:2021upt,Salazar:2019ncp,Rodriguez-Aguilar:2024efj}, quarkonia productions \cite{Czyzewski:1996bv,Kowalski:2006hc,Armesto:2014sma,Mantysaari:2020lhf,Benic:2024pqe} and so on. 
The spin dependent Pomerons: $\calP_{1T}^\perp(\kp,\delp)$ and $\calP_T(\kp,\delp)$, and the spin-dependent Odderons: $\calO_{1T}^\perp(\kp,\delp)$ and $\calO_T(\kp,\delp)$ have received more attentions after the remarkable realization that even in the eikonal approximation proton helicity-flip is possible, see \cite{Zhou:2013gsa,Boussarie:2019vmk}, and \cite{Yao:2018vcg,Hagiwara:2020mqb,Hatta:2022bxn,Agrawal:2023mzm,Benic:2024fbf,Benic:2025ral} for some phenomenological works. As we recall in Sec.~\ref{sec:GPD}, $\calP$ and $\calO_{1T}^\perp$ are the only two contributions that survive the forward limit.

The key element of the followup analysis comes from observing that the dipole GTMD contains only three complex functions \eqref{eq:Sdecompose1}: $\calS$, $\calS^\perp_{1T}$ and $\calS_{T}$, whereas the full set of leading-twist GTMDs hosts 16 independent and complex functions. By matching to the dipole GTMD we naturally anticipate relationships between the GTMDs at small-$x$. We start by expressing the elements of the helicity matrix in terms of $\calS_{\Lambda'\Lambda}$. Utilizing \eqref{eq:Wsmall} we have
\be
\begin{split}
& xU^g_{\Lambda'\Lambda} = \frac{2 N_c}{\alpha_S}\left(\kp^2 - \frac{\delp^2}{4}\right) \calS_{\Lambda'\Lambda}\,,\\
& xL^g_{\Lambda'\Lambda} = \frac{2\rmi N_c}{\alpha_S} (\kp\times\delp) \calS_{\Lambda'\Lambda}\,,\\
& xT^g_{R,\Lambda'\Lambda} = -\frac{2 N_c}{\alpha_S}\left(k_R^2 - \frac{\Delta_R^2}{4}\right) \calS_{\Lambda'\Lambda}\,.
\end{split}
\ee
For the first two lines in \eqref{eq:GTMD1} we have
\be
xH^g_{+\frac{1}{2}+1,+\frac{1}{2}+1} = xH^g_{-\frac{1}{2}+1,-\frac{1}{2}+1} = \frac{N_c}{ \alpha_S}\left[\left(\kp^2 - \frac{\delp^2}{4}\right) \calS + \rmi (\kp\times\delp)\calS\right]\,.
\ee
and so we deduce
\be
\begin{split}
& x S_{1,1a}^{0,+;g} = \frac{2 N_c}{\alpha_S}\left(\kp^2 - \frac{\delp^2}{4}\right) \calS\,,\\
& x S_{1,1b}^{0,-;g} = \frac{2 N_c}{\alpha_S} M^2 \calS\,,
\end{split}
\label{eq:Ssmallx}
\ee
while $x S_{1,1a}^{0-;g} = x S_{1,1b}^{0+;g} = 0$ in the strict eikonal limit\footnote{For small-$x$ asymptotics of the gluon helicity and the angular momentum distributions see for example \cite{Kovchegov:2017lsr} and \cite{Hatta:2016aoc,Kovchegov:2019rrz} and references therein, respectively.}. The result \eqref{eq:Ssmallx} implies that at small-$x$ the two GTMDs are {\it not} independent but related as
\be
\kp^2 x S_{1,1b}^{0,-;g} \approx M^2 x S_{1,1a}^{0,+;g}\,,
\ee
in the near-forward limit. This is the relation between the unpolarized gluon GTMD $f^g = F_{1,1}^g = S_{1,1a}^{0,+;g}$ and the spin-orbit GTMD $C_g = G_{1,1}^g = - S_{1,1b}^{0,-;g}$ that was found in \cite{Bhattacharya:2024sno}, where we also recalled the GTMD notation used in \cite{Meissner:2009ww}. It is worth noting that this result holds separately for the real and imaginary parts of the gluon GTMDs.

Next, for proton helicity-flip case we have
\be
\begin{split}
& xH^g_{+\frac{1}{2} +1,-\frac{1}{2} +1} = \frac{N_c}{\alpha_S}\left[\kp^2 - \frac{\delp^2}{4} + \rmi (\kp\times\delp)\right]\left(-\frac{k_L}{M} \calS_{1T}^\perp - \frac{\Delta_L}{M} \calS_T\right)\,,\\
& xH^g_{-\frac{1}{2} +1,+\frac{1}{2} +1} = \frac{N_c}{\alpha_S}\left[\kp^2 - \frac{\delp^2}{4} + \rmi (\kp\times\delp)\right]\left(\frac{k_R}{M} \calS_{1T}^\perp + \frac{\Delta_R}{M} \calS_T\right)\,.
\end{split}
\label{eq:Hprotflip}
\ee
Comparing with 3rd and 4th line of \eqref{eq:GTMD2} we need to express this in terms of the leading angular structures $k_{L,R}$ and $\Delta_{L,R}$. It is useful to express the scalar and the cross product in terms of $k_{L,R}$ and $\Delta_{L,R}$ as: $\kp \cdot \delp = (k_L\Delta_R + k_R \Delta_L)/2$, 
$\kp \times \delp = -\rmi (k_L\Delta_R - k_R \Delta_L)/2$. We can utilize this to come up with a suitable form of the $\epsilon$-identity\footnote{This is also known as the Schouten identity and in 2D the general form involving three vectors $\aperp$, $\bp$ and $\cperp$ is $\aperp (\bp\times\cperp) + \cperp (\aperp\times\bp) + \bp (\cperp\times\aperp) = 0$.}
\be
\begin{split}
& \rmi (\kp\times\delp) k_{L,R} = \pm(\kp \cdot \delp) k_{L,R} \mp \kp^2 \Delta_{L,R}\,,\\
& \rmi (\kp\times\delp) \Delta_{L,R} = \pm\delp^2 k_{L,R} \mp (\kp\cdot\delp) \Delta_{L,R}\,.
\end{split}
\label{eq:epsid}
\ee
With the help of \eqref{eq:epsid} we can now rewrite \eqref{eq:Hprotflip} in terms of the leading $k_{L,R}$ and $\Delta_{L,R}$ structures
\be
\begin{split}
xH^g_{+\frac{1}{2} +1,-\frac{1}{2} +1} = \frac{N_c}{\alpha_S}&\Bigg\{-\frac{k_L}{M}\left[\left(\kp^2  - \frac{\delp^2}{4}\right) \calS_{1T}^\perp + (\kp\cdot\delp)\calS_{1T}^\perp + \delp^2 \calS_T\right]\\
& - \frac{\Delta_L}{M}\left[-\kp^2 \calS_{1T}^\perp  + \left(\kp^2  - \frac{\delp^2}{4}\right)\calS_T - (\kp\cdot\delp)\calS_T\right]\Bigg\}\,,\\
xH^g_{-\frac{1}{2} +1,+\frac{1}{2} +1} = \frac{N_c}{\alpha_S}&\Bigg\{\frac{k_R}{M}\left[\left(\kp^2  - \frac{\delp^2}{4}\right)\calS_{1T}^\perp - (\kp\cdot\delp)\calS^\perp_{1T} - \delp^2 \calS_T\right]\\
& + \frac{\Delta_R}{M}\left[\kp^2 \calS_{1T}^\perp  + \left(\kp^2  - \frac{\delp^2}{4}\right)\calS_T +(\kp\cdot\delp)\calS_T\right]\Bigg\}\,.
\end{split}
\ee
Matching this onto \eqref{eq:GTMD2} we deduce that the $P^0$-wave GTMDs are given as combinations of spin-dependent dipoles
\be
\begin{split}
& xP_{1,1a}^{0,+;g} = \frac{2 N_c}{\alpha_S}\left(\kp^2  - \frac{\delp^2}{4}\right) \calS_{1T}^\perp\,,\\
& xP_{1,1a}^{0,-;g} = -\frac{2 N_c}{\alpha_S}\left[  (\kp\cdot\delp) \calS_{1T}^\perp + \delp^2 \calS_T\right]\,,\\
& xP_{1,1b}^{0,+;g} = \frac{2 N_c}{\alpha_S}\left(\kp^2  - \frac{\delp^2}{4}\right) \calS_T\,,\\
& xP_{1,1b}^{0,-;g} = \frac{2 N_c}{\alpha_S}\left[  \kp^2 \calS_{1T}^\perp +(\kp\cdot\delp) \calS_T\right]\,.
\end{split}
\label{eq:P0}
\ee
In reverse, the spin-dependent dipoles $\calS_{1T}^\perp$ and $\calS_T$ can be interpreted as small-$x$ limits of the LF-parity even $P$-wave GTMDs $P_{1,1a}^{0,+;g}$ and $P_{1,1b}^{0,+;g}$, respectively. Thus, we are lead to conclude that at small-$x$ the remaining LF-parity odd $P$-wave GTMDs can be written as linear combinations of their parity even partners. In the near-forward limit we have
\be
\begin{split}
&  \kp^2 x P^{0 -;g}_{1,1a} \approx - (\kp\cdot\delp) x P^{0 +;g}_{1,1a}\,,\\
& \kp^2 x P^{0 -;g}_{1,1b} \approx \kp^2 x P^{0 +;g}_{1,1a} + (\kp\cdot\delp) x P^{0 +;g}_{1,1b}\,.
\end{split}
\ee

Moving to the gluon helicity flip case we have
\be
xH^g_{+\frac{1}{2} +1,+\frac{1}{2} -1} = xH^g_{-\frac{1}{2} +1,-\frac{1}{2} -1} = -\frac{N_c}{\alpha_S} \left(k_R^2 - \frac{1}{4}\Delta_R^2\right) \calS\,.
\ee
Matching onto \eqref{eq:GTMD3} we find that $D$-wave GTMDs are expressed in terms of the spin independent dipole
\be
\begin{split}
& xD^{2,+;g}_{1,1a} = \frac{2 N_c}{\alpha_S} M^2 \calS\,,\\
& xD^{2,+;g}_{1,1b} = - \frac{N_c}{2\alpha_S} M^2 \calS\,.
\end{split}
\ee
while $xD'^{2,+;g}_{1,1a} = xD'^{2,+;g}_{1,1b} = 0$. We now also use \eqref{eq:Ssmallx} to connect $D$-wave and $S$-wave GTMDs at small-$x$
\be
\kp^2 x D^{2,+;g}_{1,1a} = - 4\kp^2 x D^{2,+;g}_{1,1b} = M^2 x S^{0,+;g}_{1,1a}\,.
\ee
Interesting to note here is that GTMDs with different angular momentum transfer are connected at small-$x$.

For the final case of proton and gluon helicity flips we have
\be
\begin{split}
& xH^g_{+\frac{1}{2} +1,-\frac{1}{2} -1} = \frac{N_c}{\alpha_S} \left(k_R^2 - \frac{1}{4}\Delta_R^2\right) \left(\frac{k_L}{M} \calS_{1T}^\perp + \frac{\Delta_L}{M} \calS_T\right)\,,\\
& xH^g_{-\frac{1}{2} +1,+\frac{1}{2} -1} = -\frac{N_c}{\alpha_S} \left(k_R^2 - \frac{1}{4}\Delta_R^2\right) \left(\frac{k_R}{M} \calS_{1T}^\perp + \frac{\Delta_R}{M} \calS_T\right)\,.
\end{split}
\label{eq:pgflip}
\ee
In order to match onto the first line of \eqref{eq:GTMD4} we need to write the first line in \eqref{eq:pgflip} using leading angular structure only: $k_R$ and $\Delta_R$. We can write: $k_R^2 k_L = \kp^2 k_R$, $k_R^2 \Delta_L = 2 (\kp\cdot \delp) k_R - \kp^2 \Delta_R$ and similarly when $\kp$ and $\delp$ are interchanged. We find that the $P^2$-wave GTMDs are a certain combination of spin-dependent dipoles
\be
\begin{split}
& x P^{2,+;g}_{1,1a} = -\frac{2 N_c}{\alpha_S} \left[\left(\kp^2 + \frac{\delp^2}{4}\right) \calS_{1T}^\perp + 2(\kp\cdot \delp)\calS_T\right]\,,\\
& x P^{2,+;g}_{1,1b} = \frac{2 N_c}{\alpha_S} \left[\left(\kp^2 + \frac{\delp^2}{4}\right) \calS_T + \frac{1}{2}(\kp\cdot \delp)\calS_{1T}^\perp\right]\,.
\end{split}
\label{eq:P2}
\ee
Comparing \eqref{eq:P2} and \eqref{eq:P0} we reach a result that the $P$-wave GTMDs with gluon helicity flip $\Delta S_z = 2$ can be conveniently expressed in terms of $P$-wave GTMDs with $\Delta S_z = 0$ as
\be
\begin{split}
& x P_{1,1a}^{2,+;g} \approx x P_{1,1a}^{0,+;g} - 2 x P_{1,1b}^{0,-;g}\,,\\ 
& x P_{1,1b}^{2,+;g} \approx  x P_{1,1b}^{0,+;g} - \frac{x}{2} P_{1,1a}^{0,-;g}\,, 
\end{split}
\ee
in the near forward limit.

According to the second line in \eqref{eq:GTMD4}, the second line in \eqref{eq:pgflip} needs to be rewritten in terms of $k_R^3$ and $\Delta_R^3$ and so we need to express $k_R^2 \Delta_R$ and $k_R \Delta_R^2$ in terms of $k_R^3$ and $\Delta_R^3$. Using the results in App.~\ref{sec:inner} we find
\be
\begin{split}
& k_R^2 \Delta_R = \frac{2 \delp^2 (\kp\cdot \delp)}{4 (\kp\cdot\delp)^2 - \kp^2 \delp^2} k_R^3 + \frac{\kp^4}{4(\kp\cdot\delp)^2 - \kp^2 \delp^2} \Delta_R^3\,,\\
& k_R \Delta_R^2 = \frac{\delp^4}{4(\kp\cdot\delp)^2 - \kp^2 \delp^2} k_R^3\ + \frac{2 \kp^2 (\kp\cdot \delp)}{4 (\kp\cdot\delp)^2 - \kp^2 \delp^2} \Delta_R^3\,,
\end{split}
\ee
which is also consistent with the similar decomposition shown in \cite{Bertone:2025vgy}.
Using this in the second line of \eqref{eq:pgflip} leads to
\be
\begin{split}
& x F_{1,1a}^{2,+;g} = \frac{2 N_c}{\alpha_S} M^2\left[\frac{4(\kp\cdot\delp)^2 - \kp^2 \delp^2 - \frac{\delp^4}{4}}{4(\kp\cdot\delp)^2 - \kp^2 \delp^2} \calS_{1T}^\perp + \frac{2 (\kp \cdot \delp) \delp^2}{4(\kp\cdot\delp)^2 - \kp^2 \delp^2}\calS_T \right]\,,\\
& x F_{1,1b}^{2,+;g} = -\frac{N_c}{\alpha_S} M^2\left[\frac{\kp^2(\kp\cdot\delp)}{4(\kp\cdot\delp)^2 - \kp^2 \delp^2} \calS_{1T}^\perp + \frac{1}{2}\frac{4(\kp\cdot\delp)^2 - \kp^2\delp^2 - 4\kp^4}{4(\kp\cdot\delp)^2 - \kp^2 \delp^2}\calS_T \right]\,.
\end{split}
\label{eq:Fab}
\ee
To get the forward limit, we go back to \eqref{eq:pgflip} and find
\be
x H^g_{+\frac{1}{2}+1,+\frac{1}{2}-1} = -\frac{N_c}{\alpha_S} \frac{k_R^3}{M^3} M^2 \calS_{1T}^\perp\,,
\ee
from which we extract
\be
x F^{2,+;g}_{1,1a} = \frac{2 N_c}{\alpha_S} M^2 \calS_{1T}^\perp\,,
\ee
which is consistent with taking $\delp \to 0$ in the first line of \eqref{eq:Fab}. Comparing to \eqref{eq:P0} we find a simple connection between $F$-wave and $P^0$-wave GTMD
\be
\kp^2 x F^{2,+;g}_{1,1a} \approx M^2 xP^{0,+;g}_{1,1a}.
\ee

In this section, we have obtained small-$x$ expressions for the nonvanishing leading-twist gluon GTMDs in terms of the spin-independent and spin-dependent Pomerons and Odderons. These expressions allow us to relate distributions involving parton helicity-flip to those without parton helicity-flip. In case when there is proton helicity flip we find a relation between the $D$-wave and the $S$-wave GTMDs and in case when there is proton helicity flip, we find a relation of the $P^2$-wave and $F$-wave with the $P^0$-wave GTMDs.

\subsection{The TMD and the GPD limits}
\label{sec:GPD}

We first briefly comment on the TMD limit ($\delp \to 0$) of these results, as they reproduce well-known relations from the literature. According to \eqref{eq:Sdecompose1} and \eqref{eq:Sdecompose2}, all small-$x$ TMDs should be expressed solely in terms of the distributions $\calP$ and $\calO_{1T}^\perp$. Considering the identification of the $\delp \to 0$ limit of the GTMDs \cite{Lorce:2013pza}, we have the following non-zero results for the TMDs 
\be
\begin{split}
& x f_1^g(x,\kp) = x {\rm Re}(S_{1,1a}^{0,+;g})_{\delp = 0} \to \frac{2 N_c}{\alpha_S} \kp^2 \calP(\kp,\delp = 0)\,,\\
& x h_1^g(x,\kp) = x {\rm Im}(P_{1,1a}^{2,+;g})_{\delp = 0} \to -\frac{2 N_c}{\alpha_S}\kp^2\calO_{1T}^\perp(\kp,\delp = 0)\,,\\
& x f_{1T}^{\perp g}(x,\kp) = - x {\rm Im}(P^{0,+;g}_{1,1a})_{\delp = 0} \to -\frac{2 N_c}{\alpha_S} \kp^2 \calO_{1T}^\perp(\kp,\delp = 0)\,,\\
& x h_1^{\perp g}(x,\kp)  = 2 x {\rm Re}(D^{2,+;g}_{1,1a})_{\delp = 0} \to \frac{4 N_c}{\alpha_S}M^2 \calP(\kp,\delp = 0)\,,\\
& x h_{1T}^{\perp g}(x,\kp) = 2 x {\rm Im}(F^{2,+;g}_{1,1a})_{\delp = 0} \to \frac{4 N_c}{\alpha_S} M^2 \calO_{1T}^\perp(\kp,\delp = 0)\,,
\end{split}
\label{eq:tmdl}
\ee
at small-$x$. The remaining TMDs vanish in the small-$x$ limit: $xg^g_{1L} = xh_{1L}^{\perp g} = x g_{1T}^g = 0$. The first line in \eqref{eq:tmdl} is well known in the small-$x$ computations connecting the unpolarized gluon TMD to the spin-independent Pomeron $\calP$. The gluon Sivers function $f^{\perp g}_{1T}$ becomes related to the spin-dependent Odderon $\calO^\perp_{1T}$ \cite{Zhou:2013gsa}. Eq.~\eqref{eq:tmdl} implies that the non-vanishing TMDs are related as
\be
f^g_1 = \frac{\kp^2}{2 M^2} h_1^{\perp g}\,, \qquad f_{1T}^{\perp g} = h^g_1 = -\frac{\kp^2}{2 M^2} h_{1T}^{\perp g}\,,
\ee
recovering the results found in \cite{Metz:2011wb,Boer:2015pni,Boer:2016xqr}.

Next we consider the GPD limit. We first focus on the eight leading-twist gluon GPDs which are related to the Pomerons and later address also twist-3 gluon GPDs that are related to the Odderons. At $\xi = 0$ and at small-$x$, the leading-twist GPDs are related to the real parts of GTMDs as \cite{Lorce:2013pza}:
\be
\begin{split}
 H^g & = {\rm Re}\int \rmd^2 \kp S^{0,+;g}_{1,1a}\,,\\
 E^g &= 2{\rm Re}\int \rmd^2 \kp \left[\frac{\kp\cdot\delp}{\delp^2} P^{0,+;g}_{1,1a} + P^{0,+;g}_{1,1b}\right] \,,\\
 \tilde{H}^g &= {\rm Re}\int \rmd^2 \kp S^{0,-;g}_{1,1a}\,,\\
 \xi\tilde{E}^g &= 2{\rm Re}\int \rmd^2 \kp \left[\frac{\kp\cdot\delp}{\delp^2} P^{0,-;g}_{1,1a} + P^{0,-;g}_{1,1b}\right]\,,\\
 H^g_T &= -{\rm Re}\int \rmd^2 \kp \left[\frac{\kp\cdot\delp}{\delp^2} P^{2,+;g}_{1,1a} + P^{2,+;g}_{1,1b}\right] - \frac{\delp^2}{M^2}{\rm Re}\int \rmd^2 \kp \left[\frac{4(\kp\cdot\delp)^2 -3\kp^2 \delp^2}{\delp^4}\frac{\kp\cdot\delp}{\delp^2} F^{2,+;g}_{1,1a} + F^{2,+;g}_{1,1b}\right]\,,\\
 E^g_T &= - 4{\rm Re}\int \rmd^2 \kp \left[\frac{2(\kp\cdot\delp)^2 - \kp^2 \delp^2}{\delp^4} D^{2,+;g}_{1,1a} + D^{2,+;g}_{1,1b}\right] - 8{\rm Re}\int \rmd^2 \kp \left[\frac{4(\kp\cdot\delp)^2 -3\kp^2 \delp^2}{\delp^4}\frac{\kp\cdot\delp}{\delp^2} F^{2,+;g}_{1,1a} + F^{2,+;g}_{1,1b}\right]\,,\\
\tilde{H}^g_T &= 4{\rm Re}\int \rmd^2 \kp \left[\frac{4(\kp\cdot\delp)^2 -3\kp^2 \delp^2}{\delp^4}\frac{\kp\cdot\delp}{\delp^2} F^{2,+;g}_{1,1a} + F^{2,+;g}_{1,1b}\right]\,,\\
 \tilde{E}^g_T & = - 4{\rm Re}\int \rmd^2 \kp \left[\frac{2(\kp\cdot\delp)^2 - \kp^2 \delp^2}{\delp^4} D'^{2,+;g}_{1,1a} + D'^{2,+;g}_{1,1b}\right]\,.
\end{split}
\label{eq:Gpd1}
\ee
Employing the above relations, together with those obtained in the previous section, we can now relate the GPDs to the spin-independent and spin-dependent Pomerons. For that purpose we need a Fourier expansion of the Pomeron as:
\be
\calP(\kp,\delp) = \calP_0(|\kp|,|\delp|) + \left[\frac{2(\kp\cdot\delp)^2}{\kp^2 \delp^2} - 1\right]2\calP_\epsilon(|\kp|,|\delp|) + \dots\,
\label{eq:four}
\ee
and similarly for $\calP_{1T}^\perp$ and $\calP_T$. The first term in the Fourier series \eqref{eq:four} is the isotropic Pomeron and the second term is the elliptic Pomeron, see e.~g.~\cite{Hatta:2016dxp,Zhou:2016rnt,Hagiwara:2016kam}.

In the near-forward regime the unpolarized gluon GPD $H^g$ and the helicity flip GPD $E^g$ are:
\be
\begin{split}
& xH^g = \frac{2N_c}{\alpha_S}\int \rmd^2 \kp \kp^2 \calP_0\,,\\
& x E^g = \frac{4 N_c}{\alpha_S}\int\rmd^2 \kp \kp^2 \left[\frac{\kp^2}{2M^2}(\calP_{1T,0}^\perp + \calP_{1T,\epsilon}^\perp) + \calP_{T,0}\right]\,,
\end{split}
\label{eq:gGPDs1}
\ee
The first line of this relation can be found in \cite{Hatta:2017cte}, while the second line, connecting the GPD $E^g$ with the spin-dependent Pomerons, was obtained in \cite{Hatta:2022bxn} up to the elliptic piece.
On the other hand, all gluon helicity GPDs vanish identically at small-$x$. For $\tilde{H}^g$ this is immediately obvious because the GTMD $S_{1,1a}^{0,-;g}$ vanishes at small-$x$ (see below Eq.~\eqref{eq:Ssmallx}). The GPD $\tilde{E}^g$ vanishes because under the transformation $\kp\to-\kp$, the real part of $P_{1,1a}^{0,-;g}$ is even, while the real part of $P_{1,1b}^{0,-;g}$ is odd, making the whole integrand odd under this transformation. The non-zero transversity GPDs are found to be:
\be
\begin{split}
& x H^g_T = \frac{2 N_c}{\alpha_S} \int \rmd^2 \kp \kp^2 \frac{\kp^2}{2M^2} \calP_{1T,0}^\perp \,,\\
& x E^g_T = -\frac{8 N_c}{\alpha_S} \frac{M^2}{\delp^2} \int \rmd^2 \kp \kp^2 \left[ \calP_\epsilon  + \frac{\kp^2}{ M^2}\calP^\perp_{1T,\epsilon} + 2 \calP_{T,\epsilon}\right]\,,\\
& x \tilde{H}^g_T = \frac{4N_c}{\alpha_S}\frac{M^2}{\delp^2}\int\rmd^2\kp \kp^2\left[ \frac{\kp^2}{M^2}\calP_{1T,\epsilon}^\perp + 2 \calP_{T\epsilon}\right]\,.
\end{split}
\label{eq:gGPDs2}
\ee
while the GPD $\tilde{E}_T^g$ vanishes because GTMDs $D'^{2,+;g}_{1,1a}$ and $D'^{2,+;g}_{1,1b}$ are zero at small-$x$. The connection between GPD $E_T^g$ and elliptic part of spin-independent Pomeron was already recognized in \cite{Hatta:2017cte} (the first term in the second line of \eqref{eq:gGPDs2} agrees with \cite{Hatta:2017cte} up to a factor of $-2$ due to the different conventions \cite{Diehl:2003ny} in the definition of $E^g_T$). The rest of GPD $E_T^g$, which depends on spin dependent Pomerons, together with the expressions for $H_T^g$ and $\tilde{H}_T^g$ are new. Moreover, we can combine $E^g_T$ and $\tilde{H}^g_T$ as:
\be
x E^g_T + 2 x \tilde{H}^g_T = -\frac{8N_c}{\alpha_S} \frac{M^2}{\delp^2} \int \rmd^2 \kp \kp^2 \calP_\epsilon\,,
\label{eq:gpdrel}
\ee
which emphasizes that the linear combination of $E_T^g$ and $\tilde{H}_T^g$ is, in the near-forward limit, completely controlled by the elliptic piece of spin independent Pomeron.

Whereas the Pomerons ($\calP$, $\calP_{1T}^\perp$ and $\calP_T$) contribute at the leading twist-2, the Odderons ($\calO$, $\calO_{1T}^\perp$ and $\calO_T$), on the other hand, contribute to a class of twist-3 tri-gluon GPDs. For the present purposes we only need a subset of the so-called dynamical twist-3 gluon GPDs that will be linked to the Odderons. These are constructed from the three-body gluonic operators contracted with the totally symmetric $d_{abc}$-tensor in the color space. Following Refs.~\cite{Cui:2018jha,Hatta:2024otc} we can introduce the Odderon GPDs $O_1$ and $O_2$
\be
\begin{split}
\frac{g}{P^+} \int_{-\infty}^\infty \frac{\rmd z^-}{2\pi}\frac{\rmd w^-}{2\pi} &\rme^{\rmi (x_1 + x_2)P^+ z^-} \rme^{\rmi (x_2 - x_1) P^+ w^-} d_{abc} \langle p' \Lambda' |F_a^{+i}(-z/2)F_b^{+j}(w)F_c^{+k}(z/2)|p \Lambda\rangle\\
& = \rmi \delta_{\Lambda'\Lambda}\Delta^l\left[\delta^{ik} \delta^{jl} O_1(x_1,x_2,\delp) - \delta^{ij}\delta^{kl} O_1(x_2,x_2 - x_1,\delp) - \delta^{jk}\delta^{il} O_1(x_1,x_1 - x_2,\delp)\right]\\
& + 2\rmi\Lambda\delta_{\Lambda',-\Lambda} \left(\delta_{\Lambda,-1}L^l + \delta_{\Lambda 1} R^l\right)M\\
&\times\left[\delta^{ik} \delta^{jl} O_2(x_1,x_2,\delp) + \delta^{ij} \delta^{kl} O_2(x_2,x_2 - x_1,|\delp|) + \delta^{jk}\delta^{il} O_2(x_1,x_1 - x_2,\delp)\right] + \dots\,,
\end{split}
\label{eq:odderongpd}
\ee
where the ellipsis contains sub-eikonal terms and terms with higher power in $|\delp|$. The first term was written down in complete analogy to \cite{Hatta:2024otc}, see eq. (5.1) there. The second term was adapted from its forward limit - the resulting tri-gluon PDF is familiar by its contribution to single spin asymmetries \cite{Ji:1992eu,Beppu:2010qn,Koike:2019zxc}.

We now expand the operator definition of the Odderon and match onto the tri-gluon GPDs \eqref{eq:odderongpd}, see e.~g.~\cite{Zhou:2013gsa,Benic:2024fbf,Mantysaari:2025mht} for a similar procedure. At small-$x$ we take $x_1 = x_2 \equiv x \approx 0$ and so $O_i(x_1,x_2,\delp)\approx O_i(x_2,x_2-x_1,\delp) \approx O_i(x_1,x_1 - x_2,\delp) \equiv O_i(x,\delp)$. We find the following connection between the Odderons and the GPDs $O_1$ and $O_2$
\be
\begin{split}
& O_1 = - \frac{N_c}{\pi \alpha_S}\int \rmd^2 \kp \kp^2 \frac{\kp^2}{M^2} \calO_0\,,\\
& O_2 = - \frac{N_c}{\pi \alpha_S}\int \rmd^2 \kp \kp^2 \frac{\kp^2}{M^2} \calO^\perp_{1T,0}\,,
\label{eq:oddgpd}
\end{split}
\ee
where we have kept only the isotropic parts of the Fourier expansion of the Odderon $\calO(\kp,\delp) = \calO_0(|\kp|,|\delp|) + \dots$ and likewise for $\calO_{1T}^\perp(\kp,\delp)$. In the forward limit, the connection between the spin-dependent Odderon $\calO_{1T}^\perp$ and the tri-gluon PDF $O_2(x,\delp = 0)$ was first realized in \cite{Zhou:2013gsa}.

\section{Sea-quark GTMDs at small-\ensuremath{x}}
\label{sec:seaq}

The sea-quark GTMDs are defined through the correlator \cite{Lorce:2013pza}
\be
W_{\Lambda'\Lambda}^{[\Gamma]} = \frac{1}{2}\int \frac{\rmd z^- \rmd^2 \zp}{(2\pi)^3} \rme^{\rmi x P^+ z^-} \rme^{-\rmi \kp\cdot\zp}\langle p' \Lambda' |\bar{\psi}(-z/2) \mathcal{W} \Gamma\psi(z/2)|p \Lambda\rangle\,,
\label{eq:qgtmd}
\ee
for which, in the leading-twist case, we have: $\Gamma = \gamma^+$, $\gamma^+\gamma_5$, $\rmi \sigma^{R+}\gamma_5$, $\rmi \sigma^{L+}\gamma_5$. Following the computation in \cite{Bhattacharya:2024sno}, the staple-shaped Wilson link $\mathcal{W}$ is chosen to connect the points $-z/2$ and $+z/2$ through light-cone infinity $z^- = -\infty$. Similar to the gluon case, different components of $W_{\Lambda'\Lambda}^{[\Gamma]}$ are gathered into the helicity matrix in the joint proton ($\Lambda = \pm 1/2$) and quark ($\lambda = \pm 1/2$) helicity space
\be
H^q_{\Lambda' \lambda',\Lambda \lambda} = 
\begin{pmatrix}
\frac{1}{2}(U^q + L^q)_{\Lambda'\Lambda} & \frac{1}{2}T^q_{R,\Lambda'\Lambda}\\
 \frac{1}{2}T^q_{L,\Lambda'\Lambda} & \frac{1}{2}(U^q - L^q)_{\Lambda'\Lambda}\\
\end{pmatrix}_{\lambda'\lambda}\,,
\label{eq:Hq}
\ee
where $U^q \equiv W^{[\gamma^+]}$, $L^q \equiv W^{[\gamma^+\gamma_5]}$, $T^q_R \equiv W^{[\rmi\sigma^{R+}\gamma_5]}$ and $T^q_L \equiv W^{[\rmi\sigma^{L+}\gamma_5]}$. The individual entries of $H^q_{\Lambda' \lambda',\Lambda \lambda}$ are parametrized in terms of $S$, $P$ and $D$-wave GTMDs as in \cite{Lorce:2013pza}
\be
\begin{split}
& H^q_{+\frac{1}{2} +\frac{1}{2}, +\frac{1}{2} + \frac{1}{2}} = \frac{1}{2}\left[(S^{0,+;q}_{1,1a} + S^{0,-;q}_{1,1a}) + \rmi \frac{\kp\times\delp}{M^2}(S^{0,+;q}_{1,1b} + S^{0,-;q}_{1,1b})\right]\,,\\
& H^q_{-\frac{1}{2} +\frac{1}{2}, -\frac{1}{2} + \frac{1}{2}} = \frac{1}{2}\left[(S^{0,+;q}_{1,1a} - S^{0,-;q}_{1,1a}) - \rmi \frac{\kp\times\delp}{M^2}(S^{0,+;q}_{1,1b} - S^{0,-;q}_{1,1b})\right]\,,\\
& H^q_{+\frac{1}{2} +\frac{1}{2}, -\frac{1}{2} + \frac{1}{2}} = \frac{1}{2}\left[-\frac{k_L}{M}(P^{0,+;q}_{1,1a} - P^{0,-;q}_{1,1a}) - \frac{\Delta_L}{M}(P^{0,+;q}_{1,1b} - P^{0,-;q}_{1,1b})\right]\,,\\
& H^q_{-\frac{1}{2} +\frac{1}{2}, +\frac{1}{2} + \frac{1}{2}} = \frac{1}{2}\left[\frac{k_R}{M}(P^{0,+;q}_{1,1a} + P^{0,-;q}_{1,1a}) + \frac{\Delta_R}{M}(P^{0,+;q}_{1,1b} + P^{0,-;q}_{1,1b})\right]\,,\\
& H^q_{+\frac{1}{2} +\frac{1}{2}, +\frac{1}{2} - \frac{1}{2}} = \frac{1}{2}\left[\frac{k_R}{M}(P^{1,-;q}_{1,1a} + P'^{1,-;q}_{1,1a}) + \frac{\Delta_R}{M}(P^{1,-;q}_{1,1b} + P'^{1,-;q}_{1,1b})\right]\,,\\
& H^q_{-\frac{1}{2} +\frac{1}{2}, -\frac{1}{2} - \frac{1}{2}} = \frac{1}{2}\left[\frac{k_R}{M}(P^{1,-;q}_{1,1a} - P'^{1,-;q}_{1,1a}) + \frac{\Delta_R}{M}(P^{1,-;q}_{1,1b} - P'^{1,-;q}_{1,1b})\right]\,,\\
& H^q_{+\frac{1}{2} +\frac{1}{2}, -\frac{1}{2} - \frac{1}{2}} = \frac{1}{2}\left[S^{1,-;q}_{1,1a} + \rmi \frac{\kp\times\delp}{M^2}S^{1,-;q}_{1,1b}\right]\,,\\
& H^q_{-\frac{1}{2} +\frac{1}{2}, +\frac{1}{2} - \frac{1}{2}} = -\frac{1}{2}\left[\frac{k_R^2}{M^2}D^{1,-;q}_{1,1a} + \frac{\Delta_R^2}{M^2}D^{1,-;q}_{1,1b}\right]\,.
\end{split}
\label{eq:qGTMDs}
\ee

We now calculate $H^q_{\Lambda' \lambda',\Lambda \lambda}$ in the small-$x$ framework.
Similar computations have been performed previously \cite{McLerran:1994vd,McLerran:1998nk,Mueller:1999wm,Marquet:2009ca,Kovchegov:2015zha,Xiao:2017yya,Kovchegov:2021iyc,Bhattacharya:2024sno,Bhattacharya:2025fnz,Kovchegov:2025yyl}, see also \cite{Iancu:2021rup,Hatta:2022lzj,Hatta:2024vzv,Hauksson:2024bvv} for diffractive TMDs. We follow the background propagator method \cite{McLerran:1994vd,McLerran:1998nk,Bhattacharya:2024sno,Bhattacharya:2025fnz}. Working in the covariant gauge the transverse part of the Wilson staple $\mathcal{W}$ decouples and so we can write
\be
\begin{split}
W_{\Lambda'\Lambda}^{[\Gamma]} & = \frac{1}{2}\int \rmd z^- \rmd w^- \int \rmd^2 \zp \rmd^2 \wpe \rme^{\rmi x P^+ (z^- - w^-)} \rme^{-\rmi (\kp - \delp/2)\cdot\zp}\rme^{\rmi (\kp + \delp/2)\cdot\wpe}\\
&\times\frac{1}{(2\pi)^3 \int\rmd x^-\int \rmd^2 \xp}\langle p' \Lambda' |\bar{\psi}(w) \mathcal{W}_{w,-\infty}\mathcal{W}_{-\infty,z} \Gamma \psi(z)|p \Lambda\rangle\,,
\end{split}
\ee
where $\mathcal{W}_{z,-\infty}$ is a longitudinal Wilson line extending from $-\infty$ to $z^-$. 
The fermion fields are contracted into a background propagator as $\psi(z)\bar{\psi}(w) \to - S(z,w)$ where \cite{McLerran:1994vd}
\be
\begin{split}
S(z,w) & = \theta(z^-)\theta(w^-) S_0(z,w) + \theta(z^-)\theta(-w^-) \int \rmd^4 y \delta(y^-) S_0(z,y)\gamma^- S_0(y,w)V(\yp)\\
& - \theta(-z^-)\theta(w^-) \int\rmd^4 y \delta(y^-) S_0(z,y)\gamma^- S_0(y,w)V^\dag(\yp) + \theta(-z^-)\theta(-w^-) S_0(z,w)\,,
\end{split}
\label{eq:quarkp}
\ee
and the free propagator is $S_0(z,w) = -\rmi \int_k \rme^{-\rmi k\cdot(z-w)} \slashed{k}/(k^2 + \rmi\epsilon)$.
Since the background field is localized in the light-cone $x^-$ direction, we have extended the Wilson lines to light-cone infinity $\mathcal{W}_{-\infty,z}(\zp) \to \mathcal{W}_{-\infty,\infty}(\zp) = V^\dag(\zp)$ and  $\mathcal{W}_{z,\infty}(\zp) \to \mathcal{W}_{\infty,-\infty}(\zp) = V(\zp)$, thereby connecting the sea-quark GTMDs to the gluon dipole correlator \eqref{eq:dipgtmd}.

To calculate $W_{\Lambda'\Lambda}^{[\Gamma]}$ we first split it into four terms, each corresponding to individual piece from \eqref{eq:quarkp} (see \cite{McLerran:1998nk,Bhattacharya:2025fnz})
\be
W_{\Lambda'\Lambda}^{[\Gamma]} = \sum_{i = 1}^4 W_{\Lambda'\Lambda}^{[\Gamma],i}\,.
\ee
Converting to momentum space and employing the definition of the dipole GTMD \eqref{eq:dipgtmd} the individual contributions are written as
%\be
%\langle P \Lambda |P\Lambda\rangle = 2 P^+ (2\pi)\delta(P^+ - P^+)(2\pi)^2\delta^{(2)}(\Pp - \Pp) = 2 P^+ \int_{x^-}\int_{\xp}\,.
%\ee
\be
W_{\Lambda'\Lambda}^{[\Gamma],1} = -2\pi\rmi P^+ N_c \lim_{x'\to x}\int\frac{\rmd^4 q}{(2\pi)^4}\frac{1}{q^+ - xP^+ - \rmi \epsilon}\frac{1}{q^+ - x' P^+ + \rmi \epsilon} {\rm tr}\left(\Gamma\frac{\slashed{q}}{q^2 + \rmi \epsilon}\right) \calS_{\Lambda'\Lambda}(\kp - \qp,\delp)\,,
\label{eq:W1}
\ee
\be
\begin{split}
W_{\Lambda'\Lambda}^{[\Gamma],2} &= 2\pi P^+ N_c \int\frac{\rmd^4 q_1}{(2\pi)^4}\frac{\rmd^4 q_2}{(2\pi)^4} (2\pi)\delta(q_1^- - q_2^-)(2\pi)^2\delta^{(2)}(\qtwp - \kp - \delp/2) \frac{1}{q_1^+ - xP^+ + \rmi \epsilon}\frac{1}{q_2^+ - x P^+ + \rmi \epsilon}\\
& \times{\rm tr}\left(\Gamma\frac{\slashed{q}_1}{q_1^2 + \rmi \epsilon}\gamma^- \frac{\slashed{q}_2}{q_2^2 + \rmi \epsilon}\right) \calS_{\Lambda'\Lambda}(\kp - \qonp,\delp)\,,
\end{split}
\label{eq:W2}
\ee
\be
\begin{split}
W_{\Lambda'\Lambda}^{[\Gamma],3} &= -2\pi P^+ N_c \int\frac{\rmd^4 q_1}{(2\pi)^4}\frac{\rmd^4 q_2}{(2\pi)^4} (2\pi)\delta(q_1^- - q_2^-)(2\pi)^2\delta^{(2)}(\qonp - \kp + \delp/2) \frac{1}{q_1^+ - xP^+ - \rmi \epsilon}\frac{1}{q_2^+ - x P^+ - \rmi \epsilon}\\
& \times{\rm tr}\left(\Gamma\frac{\slashed{q}_1}{q_1^2 + \rmi \epsilon}\gamma^- \frac{\slashed{q}_2}{q_2^2 + \rmi \epsilon}\right) \calS_{\Lambda'\Lambda}(\kp - \qtwp,\delp)\,,
\end{split}
\label{eq:W3}
\ee
\be
W_{\Lambda'\Lambda}^{[\Gamma],4} = -2\pi\rmi P^+ N_c \lim_{x'\to x}\int\frac{\rmd q^+}{2\pi}\frac{\rmd q^-}{2\pi} \frac{1}{q^+ - xP^+ + \rmi \epsilon}\frac{1}{q^+ - x' P^+ - \rmi \epsilon} {\rm tr}\left(\Gamma\frac{\slashed{q}}{q^2 + \rmi \epsilon}\right)_{\qp = \kp} \int\frac{\rmd^2 \kgp}{(2\pi)^2}\calS_{\Lambda'\Lambda}(\kgp,\delp)\,.
\label{eq:W4}
\ee
To get \eqref{eq:W4} we have inserted $\calS_{\Lambda'\Lambda}(\kgp,\delp)$ using \eqref{eq:sumr}.

We immediately conclude that $T^q_{R} = 0$ and $T^q_L = 0$, as the relevant Dirac traces vanish. It follows that the last four lines of \eqref{eq:qGTMDs} vanish identically, implying that each scalar function appearing there is zero:
\be
P_{1,1a}^{1,-;q} = P'^{1,-;q}_{1,1a} = P_{1,1b}^{1,-;q} = P'^{1,-;q}_{1,1b} = S_{1,1a}^{1,-;q} = S_{1,1b}^{1,-;q} = D_{1,1a}^{1,-;q} = D_{1,1b}^{1,-;q} = 0\,.
\ee
Therefore, all helicity and transversity-type quark GTMDs vanish at small-$x$. To get non-zero results, one needs to go beyond the strict eikonal limit, see \cite{Kovchegov:2017jxc,Kovchegov:2018zeq,Kovchegov:2022kyy,Santiago:2023rfl,Adamiak:2024khm} for the related computations.

We now turn to the computation of $U^q$. We take $\Gamma = \gamma^+$, for which the relevant Dirac traces are ${\rm tr}\left[\gamma^+ \slashed{q}\right] = 4 q^+$ and ${\rm tr}\left[\gamma^+ \slashed{q}_1\gamma^-\slashed{q}_2\right] = 4 (\qonp \cdot \qtwp)$. The $q^+$ integral in \eqref{eq:W1} and \eqref{eq:W4} is evaluated by contour integration. After combining the two contributions, the limit $x'\to x$ can be safely taken, yielding
\be
x W^{[\gamma^+],1}_{\Lambda'\Lambda} + x W^{[\gamma^+],4}_{\Lambda'\Lambda} = \frac{N_c}{\pi^2}\int\rmd^2 \kgp\calS_{\Lambda'\Lambda}(\kgp,\delp)\,.
\ee

For $W^{[\gamma^+],2}_{\Lambda'\Lambda}$ we compute the $q_1^+$ and $q_2^+$ integrals by contour integrations. After completing the $q_1^-$ integral we obtain
\be
x W^{[\gamma^+],2}_{\Lambda'\Lambda} = -\frac{N_c}{2\pi^2} \int\rmd^2 \qonp\rmd^2 \qtwp\delta^{(2)}\left(\qtwp - \kp - \frac{\delp}{2}\right)\frac{\qonp\cdot\qtwp}{\qonp^2 - \qtwp^2}\log\left(\frac{\qonp^2}{\qtwp^2}\right)\calS_{\Lambda'\Lambda}(\kp - \qonp, \delp)\,, 
\ee
and similarly for $W^{[\gamma^+],3}_{\Lambda'\Lambda}$. In total, the full expression is found as
\be
x U^q_{\Lambda'\Lambda}(\kp,\delp) = \frac{N_c}{2\pi^2}\int\rmd^2 \kgp\left[A_U(\kp - \kgp,\kp + \delp/2) + A_U(\kp - \delp/2,\kp - \kgp)\right]\calS_{\Lambda'\Lambda}(\kgp,\delp)\,,
\label{eq:Uq}
\ee
where $A_U(\qonp,\qtwp)$ is a hard kernel
\be
A_U(\qonp,\qtwp) = 1 - \frac{\qonp\cdot\qtwp}{\qonp^2 - \qtwp^2}\log\left(\frac{\qonp^2}{\qtwp^2}\right)\,.
\ee

In order to compute $L^q$ we take $\Gamma = \gamma^+ \gamma_5$ and it immediately follows that $W^{[\gamma^+\gamma_5],1}_{\Lambda'\Lambda} = W^{[\gamma^+\gamma_5],4}_{\Lambda'\Lambda} = 0$ since the corresponding Dirac traces vanish. Consequently, only $W^{[\gamma^+\gamma_5],2}_{\Lambda'\Lambda}$ and $W^{[\gamma^+\gamma_5],3}_{\Lambda'\Lambda}$ pieces contribute. The relevant trace is
\be
{\rm tr}\left[\gamma^+ \gamma_5\slashed{q}_1 \gamma^- \slashed{q}_2\right] = 4 \rmi \epsilon^{-+ q_1 q_2} = 4 \rmi \epsilon^{0 q_1 q_2 3} = 4\rmi \epsilon^{ij} q_1^i q_2^j = 4\rmi \qonp \times \qtwp\,.
\ee
The remaining steps of the computation proceed analogously to those for $U^q$. Thus, we can obtain the result simply by replacing $\qonp\cdot\qtwp \to -\rmi \qonp \times \qtwp$ to find
\be
x L^q_{\Lambda'\Lambda}(\kp,\delp) = -\frac{\rmi N_c}{2\pi^2}\int\rmd^2\kgp\left[A_L(\kp - \kgp,\kp + \delp/2) + A_L(\kp - \delp/2,\kp - \kgp)\right]\calS_{\Lambda'\Lambda}(\kgp,\delp)\,,
\label{eq:Lq}
\ee
with the kernel
\be
A_L(\qonp,\qtwp) = \frac{\qonp\times\qtwp}{\qonp^2 - \qtwp^2}\log\left(\frac{\qonp^2}{\qtwp^2}\right)\,.
\ee 

The results \eqref{eq:Uq} and \eqref{eq:Lq} are now master formulas from which we want to extract all the sea-quark GTMDs. The hard kernels in $U^q$ and in $L^q$ ($A_U$ and $A_L$, respectively) cases are real and independent of parton helicity. The reality of $A_U$ and $A_L$ implies that the real (imaginary) parts of the of the sea-quark GTMDs will be governed by the Pomerons (Odderons). The parton helicity independence of $A_U$ and $A_L$ implies that all GTMDs associated with parton helicity-flip will vanish since also the gluon dipole itself is independent of the gluon helicity.

According to \eqref{eq:qGTMDs}, we need to expand $U^q$ and $L^q$ to linear order in $\delp$. Since $A_U$ is symmetric with respect to $\qonp \leftrightarrow \qtwp$ exchange, we already know the hard kernel in $U^q$ must be even in $\delp \to - \delp$ and so
\be
A_U(\kp - \kgp,\kp + \delp/2) + A_U(\kp - \delp/2,\kp - \kgp) = A(\qp,\kp) + O(|\delp|^2)\,,
\ee
where we retain only the constant term
\be
A(\qp,\kp) = 2\left[1 - \frac{\qp\cdot \kp}{\kp^2 - \qp^2}\log\left(\frac{\kp^2}{\qp^2}\right)\right]\,.
\label{eq:Aconst}
\ee
Here $\qp$ is a shorthand $\qp \equiv \kp - \kgp$. The $\delp$ dependence will be obtained from the dipole GTMD $\calS_{\Lambda'\Lambda}(\kgp,\delp)$. On the other hand, $A_L$ is antisymmetric under $\qonp\leftrightarrow \qtwp$ making the hard kernel in \eqref{eq:Lq} odd in $\delp \to - \delp$ and so the expansion starts with a linear term
\be
A_L(\kp - \kgp,\kp + \delp/2) + A_L(\kp - \delp/2,\kp - \kgp) = \Delta^i B^i(\qp,\kp) + O(|\delp|^3)\,,
\label{eq:ALexp}
\ee
where
\be
\begin{split}
& B^i(\qp,\kp) = 2 \delta^{ij} k^j \frac{\qp\times\kp}{\kp^2} B_1(\qp,\kp) - \epsilon^{ij} q^j B_2(\qp,\kp)\,,\\
& B_1 = \frac{1 - \kp^2 B_2(\qp,\kp)}{\kp^2 - \qp^2}\,,\\
& B_2 = \frac{\log\left(\kp^2/\qp^2\right)}{\kp^2 - \qp^2}\,.
\end{split}
\label{eq:B12}
\ee
%After the expansion we can summarize $U^q_{\Lambda'\Lambda}$ and $L^q_{\Lambda'\Lambda}$ in the following form
%\be
%\begin{split}
%& xU^q_{\Lambda'\Lambda} = 2 N_c\int_{\kgp}A\calS_{\Lambda'\Lambda}\,,\\
%& x L^q_{\Lambda'\Lambda} = 2\rmi N_c \int_{\kgp}\left[(\kp\cdot\delp) \frac{\qp\times\kp}{\kp^2}B_1 + (\qp\times\delp) B_2\right]\calS_{\Lambda'\Lambda}\,.
%\end{split}
%\label{eq:Lusef}
%\ee
%Recalling that for this we consider $U^q$ and $L^q$ up to the linear order in $\delp$. Since the hard kernel $A$ is independent of $\delp$ this will arise from the dipole $\calS$. For $L^q$ it is the opposite - the hard kernel is already linear in $\delp$ and so we will pick up only the the forward contribution from $\calS_{\Lambda'\Lambda}$
%\be
%\calS_{\Lambda'\Lambda} \to \delta_{\Lambda\Lambda'}\calP + \rmi \delta_{\Lambda,-\Lambda'}\frac{1}{M}(\Lambda k^1 + \rmi k^2)\calO_{1T}^\perp\,.
%\ee

We start by calculating the case without proton helicity flip. Since the hard kernels $A(\qp,\kp)$ and $B^i(\qp,\kp)$ are helicity independent there will only be contributions from helicity independent Pomeron and Odderon. Up to linear order in $\delp$ we have
\be
\begin{split}
& xU^q_{+\frac{1}{2}+\frac{1}{2}} = \frac{N_c}{2\pi^2} \int\rmd^2 \kgp A \left(\calP + \rmi \frac{\kgp \cdot \delp}{M^2}\calO\right)\,,\\
& xL^q_{+\frac{1}{2}+\frac{1}{2}} = -\frac{\rmi N_c}{2\pi^2}\frac{\kp\times\delp}{\kp^2}\int\rmd^2\kgp (\qp\cdot \kp) B_2 \calP\,.
\end{split}
\label{eq:ULnoflip}
\ee
For the second term in \eqref{eq:ALexp} we have used the $\epsilon$-identity in the following form
\be
\kp^2 (\qp\times \delp) = (\qp\cdot \kp)(\kp\times\delp) + (\kp\cdot\delp)(\qp\times \kp)\,,
\ee
and also that $\int\rmd^2 \kgp (\qp\times\kp)B_{1,2}\calP = 0$.

Employing \eqref{eq:ULnoflip} in \eqref{eq:qGTMDs} we immediately deduce that the GTMDs associated with helicity distributions and orbital angular momentum are vanishing in the strict eikonal limit: $S^{0,-;q}_{1,1a} = 0$ and $S^{0,+;q}_{1,1b} = 0$. The real and the imaginary parts of $S^{0,+;q}_{1,1a}$ are parametrized as \cite{Lorce:2013pza}
\be
S^{0,+;q}_{1,1a} = X^{S0,+;q}_{1,1a} +  \rmi \frac{\kp\cdot\delp}{M^2} Y^{S0,+;q}_{1,1a}\,,
\label{eq:Sreim}
\ee
for which we find
\be
\begin{split}
& xX^{S0,+;q}_{1,1a} = \frac{N_c}{2\pi^2} \int\rmd^2\kgp A\calP_0 + \frac{N_c}{\pi^2} \left[\frac{2(\kp\cdot\delp)^2}{\kp^2 \delp^2} - 1\right] \int\rmd^2\kgp \left[\frac{2(\kp\cdot\kgp)^2}{\kp^2 \kgp^2} - 1\right] A \calP_\epsilon \,,\\
& xY^{S0,+;q}_{1,1a} = \frac{N_c}{2\pi^2}\int\rmd^2\kgp\frac{\kp\cdot\kgp}{\kp^2} A \calO_0\,.
\end{split}
\label{eq:smallxq}
\ee
We have decomposed the Pomeron into its isotropic and elliptic parts and we have kept only the isotropic part of the Odderon.
The first term in the first line in \eqref{eq:smallxq} is an off-forward generalization of a known result from \cite{Marquet:2009ca,Xiao:2017yya}. The second term in the first line takes into account the elliptic Pomeron. The second line is new, see also \cite{Kovchegov:2025yyl} for a very recent considerations covering also the case of non-zero skewness. The second line in \eqref{eq:smallxq} shows that imaginary part of unpolarized sea-quark GTMD at small-$x$ is controlled by the spin independent Odderon. As a technical aside we note that the decomposition of GTMD such as \eqref{eq:smallxq} into its real and imaginary parts is dictated by the constraints of hermiticity and light-front time reversal \cite{Lorce:2013pza}, in particular controlling whether the real or the imaginary part is even or odd under $\delp \to - \delp$ (same holds for gluons - see the discussion below Eq.~\eqref{eq:Sdecompose2}). The result that the real (imaginary) parts in \eqref{eq:smallxq} are proportional to the Pomeron (Odderon), is resulting from the fact that Pomeron and Odderon are even (odd) under LF time reversal \cite{Sievert:2014psa} and the hard kernel $A(\qp,\kp)$ is purely real.

As for $S^{0,-;q}_{1,1b}$, its real and imaginary parts are found as
\be
\begin{split}
& xX^{S0,-;q}_{1,1b} = 2N_c M^2 \int_{\kgp} (\qp\cdot\kp)B_2 \calP\,,\\
& xY^{S 0,-;q}_{1,1b} = 0\,.
\end{split}
\ee
Noticing now the relation $(\qp\cdot \kp)B_2 = 1 - \frac{1}{2}A$, we get
\be
\kp^2 xX^{S0,-;q}_{1,1b} = \frac{1}{2} M^2 xX^{S0,+;q}_{1,1a}\,.
\label{eq:spinorbitqrel}
\ee
where at $\delp \neq 0$ we have dropped the first term in the above expression thanks to the sum rule \eqref{eq:sumr}. This relation was first found in \cite{Bhattacharya:2024sno} and describes the connection between the unpolarized quark GTMD $f^q = F^q_{1,1} = X^{S0,+;q}_{1,1a}$ and the spin-orbit quark GTMD $C^q = G^q_{1,1} = - X^{S0,-;q}_{1,1b}$. Unlike for gluons, the relation between the unpolarized and the spin-orbit holds only for the real part of the GTMDs.

Next, we proceed to the case where the proton flips the helicity. For $U^q_{+\frac{1}{2} -\frac{1}{2}}$ we have
\be
x U^q_{+\frac{1}{2} -\frac{1}{2}} = - 2 N_c\int\rmd^2 \kgp A \left(\frac{k_{gL}}{M}\frac{\kgp\cdot\delp}{M^2}\calP_{1T}^\perp + \frac{\Delta_L}{M}\calP_T + \rmi \frac{k_{gL}}{M} \calO^\perp_{1T}\right)\,.
\label{eq:Uflip}
\ee
In this form, the above expression is not suitable for matching onto \eqref{eq:qGTMDs} as it needs to be expressed as a linear combination of $k_L/M$ and $\Delta_L/M$. Picking up the isotropic part of the helicity-flip Pomeron $\calP_{1T}^\perp$ in the first term of \eqref{eq:Uflip} we find
\be
\begin{split}
\int\rmd^2\kgp A \frac{k_{gL}}{M} \frac{\kgp\cdot\delp}{M^2} \calP^\perp_{1T,0} & = \frac{k_L}{M} \frac{\kp\cdot\delp}{\kp^2}\int\rmd^2\kgp A \frac{\kgp^2}{M^2} \left[\frac{2(\kp\cdot\kgp)^2}{\kp^2 \kgp^2} -1\right]\calP_{1T,0}^\perp\\
& + \frac{\Delta_L}{M} \int\rmd^2 \kgp A \frac{\kgp^2}{M^2} \left[1 - \frac{(\kp\cdot\kgp)^2}{\kp^2 \kgp^2}\right]\calP_{1T,0}^\perp\,.
\end{split}
\label{eq:isotrop}
\ee
To get this result we have performed a Fourier series decomposition of the hard coefficient $A$, followed up by angular averaging in the corresponding azimuthal angle. Details of the computation leading to \eqref{eq:isotrop} are collected in App.~\ref{sec:steps}.
Performing a similar computation for the elliptic part we obtain
\be
\begin{split}
\int\rmd^2\kgp A & \frac{k_{gL}}{M} \frac{\kgp\cdot\delp}{M^2} \left[\frac{2(\kgp\cdot\delp)^2}{\kgp^2 \delp^2} - 1\right] 2\calP^\perp_{1T,\epsilon}\\
& = \frac{k_L}{M}\frac{\kp\cdot\delp}{\kp^2}\left[\frac{2(\kp\cdot\delp)^2}{\kp^2 \delp^2} - 1\right]\int\rmd^2\kgp A  \frac{\kgp^2}{M^2}\left[\frac{8(\kp\cdot\kgp)^4}{\kp^4\kgp^4} - \frac{8(\kp\cdot\kgp)^2}{\kp^2\kgp^2} + 1\right]2 \calP^\perp_{1T,\epsilon}\\
& + \frac{\Delta_L}{M} \int\rmd^2\kgp A  \frac{\kgp^2}{M^2} \frac{2(\kp\cdot\kgp)^2}{\kp^2\kgp^2}\left[1 -\frac{(\kp\cdot\kgp)^2}{\kp^2 \kgp^2}\right]2\calP_{1T,\epsilon}^\perp\\
& + \frac{\Delta_L}{M} \left[\frac{2(\kp\cdot\delp)^2}{\kp^2 \delp^2} - 1\right]\int\rmd^2\kgp A\frac{\kgp^2}{M^2} \left[-\frac{4(\kp\cdot\kgp)^4}{\kp^4\kgp^4} + 5 \frac{(\kp\cdot\kgp)^2}{\kp^2\kgp^2} - 1\right]2 \calP^\perp_{1T,\epsilon}\,.
\end{split}
\label{eq:ellipt}
\ee
Computation steps leading to \eqref{eq:ellipt} are also given in App.~\ref{sec:steps}.

Proceeding onto the second term of \eqref{eq:Uflip} that contains the helicity-flip Pomeron $\calP_T$, its isotropic part is already proportional to $\Delta_L$. The elliptic part we write as
\be
\int\rmd^2\kgp A \left[\frac{2(\kgp\cdot\delp)^2}{\kgp^2 \delp^2}-1\right] 2\calP_{T,\epsilon} = \left[\frac{2(\kp\cdot\delp)^2}{\kp^2 \delp^2}-1\right]\int\rmd^2\kgp A \left[\frac{2(\kp\cdot\kgp)^2}{\kp^2\kgp^2}-1\right]2\calP_{T,\epsilon}\,.
\ee
In the third term of \eqref{eq:Uflip} we keep only the isotropic part of the helicity-flip Odderon $\calO_{1T}^\perp$ and this is also very simply found to be
\be
\int\rmd^2\kgp A\frac{k_{gL}}{M}\calO_{1T,0}^\perp = \frac{k_{L}}{M} \int\rmd^2\kgp A\frac{\kp\cdot\kgp}{\kp^2}\calO_{1T,0}^\perp\,.
\label{eq:O1T0U}
\ee
Altogether, the above results \eqref{eq:isotrop}-\eqref{eq:O1T0U} are inserted into \eqref{eq:Uflip} that is now written as
\be
\begin{split}
x U^q_{+\frac{1}{2} -\frac{1}{2}} & = -\frac{k_L}{M}\frac{N_c}{2\pi^2} \int\rmd^2\kgp A\left\{\frac{\kgp^2}{\kp^2}\left[\frac{2(\kp\cdot\kgp)^2}{\kp^2\kgp^2} - 1\right]\frac{\kp\cdot\delp}{M^2}\calP^\perp_{1T,0} + \rmi \frac{\kp\cdot\kgp}{\kp^2}\calO_{1T,0}^\perp\right\}\\
& - \frac{k_L}{M}\frac{\kp\cdot\delp}{M^2}\left[\frac{2(\kp\cdot\delp)^2}{\kp^2\delp^2} - 1\right]\frac{N_c}{2\pi^2}\int\rmd^2\kgp A\frac{\kgp^2}{\kp^2} \left[\frac{8(\kp\cdot\kgp)^4}{\kp^4\kgp^4} - \frac{8(\kp\cdot\kgp)^2}{\kp^2\kgp^2} + 1\right] 2\calP^\perp_{1T,\epsilon}\Bigg\}\\
& - \frac{\Delta_L}{M}\frac{N_c}{2\pi^2}\int\rmd^2 \kgp A\left\{\frac{\kgp^2}{M^2}\left[1 - \frac{(\kp\cdot\kgp)^2}{\kp^2\kgp^2}\right]  \left[\calP^\perp_{1T,0} +  \frac{2(\kp\cdot\kgp)^2}{\kp^2\kgp^2}2\calP_{1T,\epsilon}^\perp\right] + \calP_{T,0}\right\}\\
& - \frac{\Delta_L}{M}\left[\frac{2(\kp\cdot\delp)^2}{\kp^2\delp^2} - 1\right]\frac{N_c}{2\pi^2}\int\rmd^2\kgp A \\
&\times\left\{\frac{\kgp^2}{M^2} \left[-\frac{4(\kp\cdot\kgp)^4}{\kp^4\kgp^4} + \frac{5 (\kp\cdot\kgp)^2}{\kp^2\kgp^2} - 1\right]2 \calP^\perp_{1T,\epsilon} + \left[\frac{2(\kp\cdot\kgp)^2}{\kp^2\kgp^2} - 1\right] 2 \calP_{T,\epsilon}\right\}\,.
\end{split}
\label{eq:Uflip2}
\ee

According to Eqs.~\eqref{eq:Lq}, \eqref{eq:ALexp} and \eqref{eq:B12}, the proton helicity-flip component of $L^q$ reads
\be
x L^q_{+\frac{1}{2}-\frac{1}{2}} = \frac{N_c}{2\pi^2} \int\rmd^2\kgp \left[2(\kp\cdot\delp) \frac{\qp \times \kp}{\kp^2} B_1 + (\qp\times\delp) B_2\right]\frac{k_{gL}}{M} \calO^\perp_{1T,0}\,.
\label{eq:Lflip}
\ee
Performing a computation that expresses $L^q_{+\frac{1}{2} -\frac{1}{2}}$ as a linear combination of $k_L$ and $\Delta_L$ (the details are in the App.~\ref{sec:steps}) we find
\be
\begin{split}
x L^q_{+\frac{1}{2}-\frac{1}{2}} & = \frac{k_L}{M}\frac{\kp\cdot\delp}{\kp^2} 2 \rmi N_c\int_{\kgp} \left[\left(\frac{(\kp\cdot\kgp)^2}{\kp^2\kgp^2} -1\right) 2 B_1 + \left(\frac{2(\kp\cdot\kgp)^2}{\kp^2 \kgp^2} - \frac{\kp\cdot\kgp}{\kgp^2} - 1\right)B_2\right]\kgp^2\calO_{1T,0}^\perp\\
& + \frac{\Delta_L}{M}2\rmi N_c\int_{\kgp} B_2\left[-\frac{(\kp\cdot\kgp)^2}{\kp^2\kgp^2} + \frac{\kp\cdot\kgp}{\kgp^2}\right] \kgp^2\calO^\perp_{1T,0}\,.
\end{split}
\label{eq:Lflip2}
\ee

Eqs.~\eqref{eq:Uflip2} and \eqref{eq:Lflip2} are now in a suitable form for the extraction of the relevant GTMDs via \eqref{eq:qGTMDs}. We get $U^q_{-\frac{1}{2}+\frac{1}{2}}$ and $L^q_{-\frac{1}{2}+\frac{1}{2}}$ from \eqref{eq:Uflip2} and \eqref{eq:Lflip2} by LF parity transformation, that is replacing $k_L \to - k_L$ and $\Delta_L \to - \Delta_R$ (with overall sign flip in $L^q$ due to $\gamma_5$). This step completes the helicity matrix \eqref{eq:Hq}. Parametrizing the real and imaginary parts of the sea-quark GTMDs as \cite{Lorce:2013pza} 
\be
\begin{split}
& P^{0,+;q}_{1,1a} = \frac{\kp\cdot\delp}{M^2}X^{P0,+;q}_{1,1a} + \rmi Y^{P0,+;q}_{1,1a}\,,\\
& P^{0,-;q}_{1,1a} = X^{P0,-;q}_{1,1a} + \rmi \frac{\kp\cdot\delp}{M^2}Y^{P0,-;q}_{1,1a}\,,\\
& P^{0,+;q}_{1,1b} = X^{P0,+;q}_{1,1b} + \rmi \frac{\kp\cdot\delp}{M^2}Y^{P0,+;q}_{1,1b}\,,\\
& P^{0,-;q}_{1,1b} = \frac{\kp\cdot\delp}{M^2}X^{P0,-;q}_{1,1b} + \rmi Y^{P0,-;q}_{1,1b}\,,
\end{split}
\ee
we find
\be
\begin{split}
x X_{1,1a}^{P0,+;q} & = \frac{N_c}{2\pi^2}\int\rmd^2\kgp A \frac{\kgp^2}{\kp^2}\left[\frac{2(\kp\cdot\kgp)^2}{\kp^2\kgp^2} - 1\right]\calP^\perp_{1T,0}\\
& + \left[\frac{2(\kp\cdot\delp)^2}{\kp^2\delp^2} - 1\right]\frac{N_c}{\pi^2}\int\rmd^2\kgp A \frac{\kgp^2}{\kp^2}\left[\frac{8(\kp\cdot\kgp)^4}{\kp^4\kgp^4} - \frac{8(\kp\cdot\kgp)^2}{\kp^2\kgp^2} + 1\right] \calP^\perp_{1T,\epsilon}\,,\\
x Y_{1,1a}^{P0,+;q} & = \frac{N_c}{2\pi^2}\int\rmd^2\kgp A\frac{\kp\cdot\kgp}{\kp^2}\calO^\perp_{1T,0}\,,\\
x Y_{1,1a}^{P0,-;q} & = \frac{N_c}{2\pi^2}\frac{M^2}{\kp^2}\int\rmd^2\kgp \left[\left(\frac{(\kp\cdot\kgp)^2}{\kp^2\kgp^2} -1\right) 2 B_1 + \left(\frac{2(\kp\cdot\kgp)^2}{\kp^2 \kgp^2} - \frac{\kp\cdot\kgp}{\kgp^2} - 1\right)B_2\right]\kgp^2\calO^\perp_{1T,0}\,,\\
x X_{1,1b}^{P0,+;q} & = \frac{N_c}{2\pi^2}\int\rmd^2\kgp A\left\{\frac{\kgp^2}{M^2}\left[1 - \frac{(\kp\cdot\kgp)^2}{\kp^2\kgp^2}\right]\left[\calP^\perp_{1T,0} + \frac{2(\kp\cdot\kgp)^2}{\kp^2\kgp^2} 2 \calP_{1T,\epsilon}^\perp\right] + \calP_{T,0}\right\}\\
& + \left[\frac{2(\kp\cdot\delp)^2}{\kp^2\delp^2} - 1\right] \frac{N_c}{\pi^2}\int\rmd^2\kgp A\left\{\frac{\kgp^2}{M^2}\left[ - \frac{4(\kp\cdot\kgp)^4}{\kp^4\kgp^4} + \frac{5 (\kp\cdot\kgp)^2}{\kp^2\kgp^2} - 1\right]\calP_{1T,\epsilon}^\perp + \left[\frac{2(\kp\cdot\kgp)^2}{\kp^2\kgp^2} - 1\right] \calP_{T,\epsilon}\right\}\,,\\
xY_{1,1b}^{P0,-;q} & = \frac{N_c}{2\pi^2} \int\rmd^2\kgp B_2\left[-\frac{(\kp\cdot\kgp)^2}{\kp^2\kgp^2} + \frac{\kp\cdot\kgp}{\kgp^2}\right]\kgp^2\calO^\perp_{1T,0}\,,
\end{split}
\label{eq:qgtmd2}
\ee
while $X_{1,1a}^{P0,-;q} = Y_{1,1b}^{P0,+;q} = X_{1,1b}^{P0,-;q} = 0$.
%{
%\color{orange}Under the time-reversal transformation, the real parts of quark GTMDs are even \cite{Lorce:2013pza}, so it is no surprise that they are described in terms of T-even Pomerons \cite{Kovchegov:2012ga}. On the other hand, the T-odd imaginary parts of GTMDs are governed by the T-odd Odderons. 
%Although this goes beyond the scope of this work, the obtained expressions for the sea-quark GTMDs can be further calculated numerically by implementing models for the Pomerons and Odderons. For example... 
%}
As expected, in the real part of the sea-quark GTMDs, the isotropic part of the GTMDs is governed by the isotropic Pomerons, and the elliptic part by the elliptic Pomerons. As a single exception we find that in $X_{1,1b}^{P0,+;q}$ the isotropic part has a term that is also proportional to $\calP_{1T,\epsilon}^\perp$.
%Using these general expressions for the sea-quark GTMDs: \eqref{eq:smallxq} and \eqref{eq:qgtmd2}, one can perform a numerical computation from various models of the Pomerons and the Odderons \cite{Benic:2025ral}, including their helicity dependence \cite{}. Their small-$x$ behaviors of can likewise be studied by solving the BK/BFKL evolution equations.

\subsection{Perturbative tails of the sea-quark GTMDs at small-\ensuremath{x}}

From the obtained sea-quark GTMD expressions \eqref{eq:smallxq} and \eqref{eq:qgtmd2} we can deduce their perturbative, or high-$\kp$, tails. We naturally expect that the remaining $\kgp$ integrations factorizes into the gluon GPDs that were given in Sec.~\ref{sec:GPD}. This serves not only as an additional insight into the above general results, but also as a consistency check for the factorization. As a specific example, the high-$\kp$ tail of $S$-wave sea-quark GTMD in \eqref{eq:smallxq} is obtained by expanding $A$ through $\kp^2 \gg \kgp^2$ as
\be
A \to \frac{-2 (\kp\cdot\kgp)^2 + 3\kp^2 \kgp^2}{3\kp^4} + \frac{\kp\cdot\kgp}{\kp^2}\frac{-4(\kp\cdot\kgp)^2 + 5 \kp^2 \kgp^2 }{3\kp^4}\,.
\label{eq:Aexp}
\ee
Focusing on the real part, we insert the $\sim \kgp^2$ term from \eqref{eq:Aexp} into the first line of \eqref{eq:smallxq} and get
\be
X_{1,1a}^{S0,+;q}(x,\kp,\delp) = \frac{\alpha_S}{6\pi^2} \frac{H^g(x,\delp)}{\kp^2} + \frac{2(\kp\cdot\delp)^2 - \kp^2 \delp^2}{\kp^2 M^2}\frac{\alpha_S}{48 \pi^2}\frac{E^g_T(x,\delp) + 2 \tilde{H}^g_T(x,\delp)}{\kp^2}\,.
\label{eq:Sqtail}
\ee
We have used \eqref{eq:gGPDs1} in the first term and \eqref{eq:gpdrel} in the second term of the above expression. The forward limit of this formula establishes the known small-$x$ result \cite{McLerran:1998nk} for the $\kp$-tail of the sea-quark TMD $f_1^q = X_{1,1a}^{S0,+;q}|_{\delp = 0}$ in terms of the gluon PDF $f_1^g(x)$
\be
f_1^q(x,\kp) = \frac{\alpha_S}{6\pi^2} \frac{f_1^g(x)}{\kp^2}\,.
\ee
The second term in \eqref{eq:Sqtail} is new and represents an off-forward correction in the small-$x$ limit. The combination $E^g_T + 2 \tilde{H}^g_T$ also appears in the computation in \cite{Bertone:2025vgy}, see their (33). The imaginary part of the $S$-wave GTMD contains the same hard coefficient, but is controlled by the Odderon, which is odd in $\kgp \to - \kgp$. For this symmetry reason we will pick up the cubic term from the expansion \eqref{eq:Aexp} and find
\be
Y^{S0,+,q}_{1,1a}(x,\kp,\delp) = -\frac{\alpha_S}{6\pi} \frac{M^2}{\kp^2}\frac{1}{x}\frac{O_1(x,\delp)}{\kp^2}\,.
\ee
We have used \eqref{eq:oddgpd} that connects the spin-independent Odderon $\calO_0$ with the corresponding spin-independent twist-3 GPD $O_1$ from \eqref{eq:odderongpd}.

The high-$\kp$ tail of the real part of the sea-quark spin-orbit GTMD $S^{0,-;q}_{1,1b}$ is completely determined by the relationship to the unpolarized sea-quark GTMD \eqref{eq:spinorbitqrel} where the obtained result agrees with \cite{Kanazawa:2014nha} in the forward limit - their Eq.~(101) matches our result after the $z$-integration is performed in accordance with the small-$x$ limit. For the off-forward correction we find the same GPD combination $E^{g}_T + 2\tilde{H}^g_T$, which in our case is quadratic and traceless in $\delp$ \eqref{eq:spinorbitqrel}, while in Eq.~(101) in \cite{Kanazawa:2014nha} it is given by $\delp^2$, without angular dependence. We were not able, however, to find agreement with the result in \cite{Bertone:2025vgy}, for this particular GTMD: their Eq.~(38) does not contain the $H^g$ piece that is in \eqref{eq:Sqtail} and so must be in $S^{0,-;q}_{1,1b}$ via \eqref{eq:spinorbitqrel} (and is found in Eq.~(101) of \cite{Kanazawa:2014nha}). The $E^g_T + 2 \tilde{H}^g_T$ piece in \cite{Bertone:2025vgy} is linear in $\delp$, while our result \eqref{eq:Sqtail} (and \cite{Kanazawa:2014nha}) gives $\sim \delp^2$ dependence.

Moving onto the $P$-wave sea-quark GTMDs that are associated with proton helicity flips, we apply the similar procedure as outlined above. The high-$\kp$ tails of $X_{1,1a}^{P0,+;q}$ and $Y_{1,1a}^{P0,+;q}$ are then found as
\be
\begin{split}
& X^{P0,+;q}_{1,1a}(x,\kp,\delp) = - \frac{\alpha_S}{12\pi^2} \frac{M^2}{\kp^2}\frac{H_T^g(x,\delp)}{\kp^2}\,,\\
& Y^{P0,+;q}_{1,1a}(x,\kp,\delp) = - \frac{\alpha_S}{6\pi} \frac{M^2}{\kp^2}\frac{1}{x}\frac{O_2(x,\delp)}{\kp^2}\,.
\end{split}
\label{eq:Pqhigh}
\ee
To get the first line of the above expression we have used the connection between the isotropic part of the helicity-flip Pomeron $\calP_{1T,0}^\perp$ and the gluon GPD $H_T^g$, see the first line in \eqref{eq:gGPDs2}. This contribution can also be identified in \cite{Bertone:2025vgy} by taking the small-$x$ limit of their (35). Notice that the elliptic part from $X^{P0,+;q}_{1,1a}$ in \eqref{eq:qgtmd2} has decoupled in the high-$\kp$ limit. To get the second line of the above expression we have used the second line in \eqref{eq:oddgpd} that establishes the connection between the spin dependent Odderon and the corresponding twist-3 gluon GPD $O_2$ from \eqref{eq:odderongpd}. It is worthwhile to mention that $Y^{P0,+}_{1,1a}$ is the only $P$-wave type sea-quark GTMD with a forward limit at small-$x$, where it defines the sea-quark Sivers function: $f^{\perp,q}_{1T} = - Y^{P0,+}_{1,1a}|_{\delp = 0}$ \cite{Lorce:2013pza}. The forward limit result agrees with the known results from the literature \cite{Zhou:2013gsa,Dong:2018wsp} both through the general connection in \eqref{eq:qgtmd2} as well as in the high-$\kp$ limit in \eqref{eq:Pqhigh}. 

To get the tail of the GTMD $Y^{P0,-;q}_{1,1a}$ we make a high-$\kp$ expansion of $B_1$ and $B_2$ that were introduced in \eqref{eq:B12}. The expansion itself gives a lengthy expression that we do not reproduce here. The odd powers of $\kgp$ vanish after the angular averaging. This leaves out the $\kgp^4$ term for which we again use \eqref{eq:oddgpd} in order to end up with
\be
Y^{P0,-;q}_{1,1a}(x,\kp,\delp) = \frac{\alpha_S}{12\pi} \frac{M^4}{\kp^4} \frac{1}{x}\frac{O_2(x,\delp)}{\kp^2}\,.
\ee

The tail of the spin-flip GTMD $X_{1,1b}^{P0,+;q}$ is found to be
\be
X_{1,1b}^{P0,+;q}(x,\kp,\delp) = \frac{\alpha_S}{24\pi^2} \frac{H_T^g(x,\delp) + 2 E^g(x,\delp)}{\kp^2} - \frac{2(\kp\cdot\delp)^2 - \kp^2\delp^2}{\kp^2 M^2}\frac{\alpha_S}{48\pi^2}\frac{\tilde{H}^g_T(x,\delp)}{\kp^2}\,.
\ee
To get this result for the isotropic part $X_{1,1b}^{P0,+;q}$ we have found that the resulting linear combination of $\calP_{1T,0}^\perp$, $\calP_{T,0}$ and $\calP_{1T,\epsilon}^\perp$ is expressed through the gluon GPDs $H_T^g$ (first line in \eqref{eq:gGPDs2}) and $E^g$ (second line in \eqref{eq:gGPDs1}). In the elliptic part, the combination of elliptic Pomerons $\calP_{1T,\epsilon}^\perp$ and $\calP_{T,\epsilon}$ precisely matches the linear combination found in the GPD $\tilde{H}^g_T$ - see the last line in \eqref{eq:gGPDs2}. These results are again in agreement with \cite{Bertone:2025vgy}, see their (36).
 
For the tail of the final GTMD, $Y_{1,1b}^{P0,-;q}$, we again expand the hard part and find
\be
Y_{1,1b}^{P0,-;q}(x,\kp,\delp) = \frac{\alpha_S}{12\pi}\frac{M^2}{\kp^2}\frac{1}{x}\frac{O_2(x,\delp)}{\kp^2}\,.
\ee

The general expressions for the imaginary part of the $P$-wave sea-quark GTMDs from \eqref{eq:qgtmd2} are all rather different, however, thanks to the fact that they are all governed by the helicity-flip Odderon $\calO_{1T}^\perp$, their high-$\kp$ tails become related in a very simple way as
\be
\kp^2 Y^{P0,+;q}_{1,1a} = -2 M^2Y^{P0,-;q}_{1,1a} = - 2\kp^2 Y^{P0,+;q}_{1,1b}\,.
\ee

\subsection{The GPD limit}
We close this section by considering the GPD limit of sea-quark GTMDs. At $\xi = 0$ and in the small-$x$ limit, leading-twist quark GPDs are related to the real part of GTMDs as \cite{Lorce:2013pza}:
\be
\begin{split}
 H^q & = {\rm Re}\int \rmd^2 \kp S^{0,+;q}_{1,1a}\,,\\
 E^q &= 2{\rm Re}\int \rmd^2 \kp \left[\frac{\kp\cdot\delp}{\delp^2} P^{0,+;q}_{1,1a} + P^{0,+;q}_{1,1b}\right] \,,\\
 \tilde{H}^q &= {\rm Re}\int \rmd^2 \kp S^{0,-;q}_{1,1a}\,,\\
 \xi\tilde{E}^q &= 2{\rm Re}\int \rmd^2 \kp \left[\frac{\kp\cdot\delp}{\delp^2} P^{0,-;q}_{1,1a} + P^{0,-;q}_{1,1b}\right]\,,\\
 H^q_T &= \frac{1}{2}{\rm Re}\int \rmd^2 \kp S^{1,-;q}_{1,1a} + \frac{\delp^2}{2M^2}{\rm Re}\int \rmd^2 \kp \left[\frac{2(\kp\cdot\delp)^2 -\kp^2 \delp^2}{\delp^4} D^{1,-;q}_{1,1a} + D^{1,-;q}_{1,1b}\right]\,,\\
 E^q_T &= 2{\rm Re}\int \rmd^2 \kp \left[\frac{\kp \cdot \delp}{\delp^2} P^{1,-;q}_{1,1a} + P^{1,-;q}_{1,1b}\right] +4{\rm Re}\int \rmd^2 \kp \left[\frac{2(\kp\cdot\delp)^2 -\kp^2 \delp^2}{\delp^4} D^{1,-;q}_{1,1a} + D^{1,-;q}_{1,1b}\right]\,,\\
\tilde{H}^q_T &= -2{\rm Re}\int \rmd^2 \kp \left[\frac{2(\kp\cdot\delp)^2 -\kp^2 \delp^2}{\delp^4} D^{1,-;q}_{1,1a} + D^{1,-;q}_{1,1b}\right]\,,\\
 \tilde{E}^q_T & = 2{\rm Re}\int \rmd^2 \kp \left[\frac{\kp\cdot\delp}{\delp^2} P'^{1,-;q}_{1,1a} + P'^{1,-;q}_{1,1b}\right] \,.
\end{split}
\label{eq:Gpdquarks}
\ee
Since only the GTMDs $S_{1,1a}^{0,+;q}$, $S_{1,1b}^{0,-;q}$, $P_{1,1a}^{0,+;q}$ and $P_{1,1b}^{0,+;q}$ posses non-zero real part only the GPDs $H^q$ and $E^q$ are non-vanishing for which we obtain
\be
\begin{split}
    &xH^q = \frac{N_c}{2\pi^2} \int \rmd^2 \kp \rmd^2 \kgp A \mathcal{P}_0\\
    &xE^q = \frac{N_c}{\pi^2} \int \rmd^2\kp \rmd^2 \kgp A\left[\frac{\kgp^2}{2M^2}\big(\mathcal{P}_{1T,0}^\perp + \mathcal{P}_{1T,\epsilon}^\perp\big) + \mathcal{P}_{T,0}\right]\,.
\end{split}
\ee
Interestingly, the resulting Pomeron decomposition in the sea-quark GPDs is identical to that of the analogous gluon GPDs - see \eqref{eq:gGPDs1}. The only difference is that, in the quark case, the kernel $A$ appears, whereas in the gluon case it is replaced by $\kgp^2$. The above result can be useful for calculations of the small-$x$ contributions to the proton form-factors \cite{Hagiwara:2024wqz}.

\section{Conclusions}
\label{sec:concl}

We have performed a systematic computation of all gluon and sea-quark GTMDs at small-$x$ and vanishing skewness $\xi$ in terms of the (helicity-dependent) gluon dipole distribution $\calS_{\Lambda'\Lambda}(\kp,\delp)$ given by \eqref{eq:dipgtmd}. We find that the small-$x$ limit connects gluon GTMDs with different orbital angular momentum transfer or helicity transfer between gluons. This is akin to the connection between the unpolarized and the linearly polarized gluon TMD \cite{Metz:2011wb}. Out of the total of 16 gluon GTMDs, the four helicity-type GTMDs vanish in the small-$x$ limit, and the remaining 12 are fully described in terms of the three Pomerons and Odderons - the real and the imaginary parts of $\calS_{\Lambda'\Lambda}(\kp,\delp)$. In other words, within the remaining set of non-vanishing 12 gluon GTMDs only three are independent at small-$x$. As a corollary, we have clarified the role of Pomerons and Odderons in the gluon GPDs, leading to new results for transversity-type gluon GTMDs.

For the sea-quark GTMDs, we find that all helicity-type and transversity-type GTMDs vanish, so in the full set of 16 sea-quark GTMDs the non-zero entries concern only the total of 6 GTMDs; the unpolarized GTMD \cite{Bhattacharya:2025fnz}, the spin-orbit GTMD \cite{Bhattacharya:2024sno} and the GTMDs for transversely polarized proton. The closed-form analytic expressions are determined by the gluon dipole $\calS_{\Lambda'\Lambda}(\kp,\delp)$ convoluted with a hard kernel that is real (and independent of parton helicity). For this reason, the real (imaginary) parts of all the sea-quark GTMDs are controlled by the real (imaginary) part of $\calS_{\Lambda'\Lambda}(\kp,\delp)$ - the Pomerons (Odderons). In particular, the imaginary part of the unpolarized sea-quark GTMD is determined by the spin-independent Odderon and the imaginary parts of the proton helicity-flip $P$-wave GTMDs are given in terms of the spin-dependent Odderon. Each GTMD is controlled by its own hard kernel and in the general kinematics we find no new connections between the sea-quark GTMD apart from the known unpolarized and the spin-orbit GTMD \cite{Bhattacharya:2024sno} connection. In the perturbative, high-$\kp$, limit we were able to factorize all the sea-quark GTMDs in terms of the small-$x$ gluon GPDs. The real parts of the GTMDs are all sourced by different gluon GPDs and therefore unrelated even at high-$\kp$. The imaginary part of the three non-zero $P$-wave GTMDs are all determined by the spin-dependent tri-gluon GPD, and so they become related at high-$\kp$ where the hard kernels are simple. 

The obtained relations are useful as universal constraints that can be used to simplify parametrizations of the GTMDs for the future explorations at the EIC. The analytical formulas can also be used in practice as a starting point for numerical computations of all the GTMDs using models of Pomerons and Odderons (for ex.~\cite{Dumitru:2018vpr,Dumitru:2019qec,Dumitru:2023sjd,Benic:2025ral}) as an initial condition for their small-$x$ \cite{Marquet:2009ca} evolutions or Sudakov evolutions \cite{Xiao:2017yya,Boer:2022njw} that becomes a foundation for further phenomenological studies. Other possible extensions concern calculations of the Weisz\" acker-Williams type gluon GTMDs \cite{Dominguez:2011wm} that is related to the gluon quadrupole operator, the diffractive GTMDs \cite{Iancu:2021rup,Hatta:2022lzj,Hatta:2024vzv,Hauksson:2024bvv,Bhattacharya:2025fnz}, skewness corrections \cite{Bhattacharya:2025fnz,Kovchegov:2025yyl} and so on.

\begin{acknowledgments}
We thank Yoshitaka Hatta for his notes on background propagator method. This work was supported by the Croatian Science Foundation (HRZZ) no. 5332 (UIP-2019-04).
\end{acknowledgments}

\appendix
\section{Parametrization and Inner Products in Two Dimensions}
\label{sec:inner}

The general structure of the parametrization of $H_{\Lambda'\lambda',\Lambda \lambda}^{g,q}$, which accounts for the change in the orbital angular momentum, is determined by terms of the form
\begin{align}\label{pocetni}
     k_R^N k_L^n \Delta_R^M \Delta_L^m\,,
\end{align}
characterized by the “weight” $W=N-n+M-m$. Without loss of generality, in this analysis we take $W>0$. Since $k_L$ and $\Delta_L$ can be expressed in terms of the $k_R$ and $\Delta_R$ through the relations $k_L=\kp^2/k_R$ and $\Delta_L = \delp^2/\Delta_R$, we choose $k_R^{W}$ and $\Delta_R^{W}$ as independent structures\footnote{In the case $W<0$ independent structures would instead be $k_L^{|W|}$ and $\Delta_L^{|W|}$.}. to which (\ref{pocetni}) can be decomposed as
\begin{align}\label{dekomp}
    k_R^N k_L^n \Delta_R^M \Delta_L^m &= a k_R^W + b \Delta_R^W\,.
\end{align}
To determine the coefficients $a$ and $b$ systematically, we define an inner product for two vectors of the same “weight” $p$ as
\begin{align}\label{definicija}    
    \langle v^p_R , w^p_R \rangle = \frac{1}{2} (v^p_R w^p_L + v^p_L w^p_R)\,.
\end{align}
This inner product is symmetric and linear in each argument
\begin{align}
    \langle v^p_R , a w^p_R + b z_R^p \rangle = a \langle v^p_R , w^p_R \rangle + b\langle v^p_R ,  z_R^p \rangle \ \ \ (a, b \in \mathbb{R}).
\end{align}
Furthermore, it satisfies
\be
    \langle k_R^p \Delta_R^q, k_R^{p+q} \rangle 
    = (k_R k_L)^p \langle \Delta_R^q, k_R^q \rangle = \langle k_R, k_R \rangle^p \langle \Delta_R^q, k_R^q \rangle\,,
\label{A5}
\ee
as well as analogous relations for other combinations, such as
\be
\langle k_R^p \Delta_L^q, v_R^{p-q} \rangle = \frac{1}{2}(k_R^p \Delta_L^q v_L^{p-q} + k_L^p \Delta_R^q v_R^{p-q}) = \langle k_R^p , \Delta_R^q v_R^{p-q} \rangle\,.
\label{A6}
\ee

Writing the vectors in (\ref{definicija}) in polar form, we obtain
\be
\cos(p\phi_{k\Delta}) = \frac{\langle \Delta_R^p,k_R^p\rangle}{(\sqrt{\langle\Delta_R,\Delta_R\rangle\langle k_R,k_R\rangle})^p} = T_p\Bigg(\frac{\langle \Delta_R,k_R\rangle}{(\sqrt{\langle\Delta_R,\Delta_R\rangle\langle k_R,k_R\rangle})}\Bigg)\,,
\label{angle}
\ee
where the norm is given by $\langle k_R^p, k_R^p \rangle =  ( k_R k_L )^p = \langle k_R, k_R \rangle^p$.
In the second line we have used the definition of the Chebyshev polynomials of the first kind. This expression allows products of higher powers to be evaluated in a relatively simple way.  

Returning to (\ref{dekomp}) and projecting out the coefficients, we have
\begin{align}    
    \langle k_R^N k_L^n \Delta_R^M \Delta_L^m, k_R^W\rangle &= a \langle k_R^W, k_R^W \rangle + b \langle k_R^W, \Delta_R^W \rangle\,, \\
 \langle k_R^N k_L^n \Delta_R^M \Delta_L^m, \Delta_R^W\rangle &= a \langle \Delta_R^W, k_R^W \rangle + b \langle \Delta_R^W, \Delta_R^W \rangle\,,
\end{align}
from which follows that $a$ and $b$ are given as
\begin{align}
    a &= \frac{\langle k_R^N k_L^n \Delta_R^M \Delta_L^m, k_R^W\rangle \langle \Delta_R^W, \Delta_R^W \rangle - \langle k_R^N k_L^n \Delta_R^M \Delta_L^m, \Delta_R^W\rangle  \langle k_R^W, \Delta_R^W \rangle }{\langle k_R^W, k_R^W \rangle  \langle \Delta_R^W, \Delta_R^W \rangle - \langle k_R^W, \Delta_R^W \rangle^2 }\,, \\
    b &=\frac{-\langle k_R^N k_L^n \Delta_R^M \Delta_L^m, k_R^W\rangle  \langle k_R^W, \Delta_R^W \rangle + \langle k_R^N k_L^n \Delta_R^M \Delta_L^m, \Delta_R^W\rangle \langle k_R^W, k_R^W \rangle }{\langle k_R^W, k_R^W \rangle  \langle \Delta_R^W, \Delta_R^W \rangle - \langle k_R^W, \Delta_R^W \rangle^2 }\,.
\end{align}

The inner products $\langle k_R^N k_L^n \Delta_R^M \Delta_L^m, k_R^W\rangle$ and $\langle k_R^N k_L^n \Delta_R^M \Delta_L^m, \Delta_R^W\rangle$ can be expressed in terms of $\langle k_R, k_R\rangle,\langle \Delta_R, \Delta_R\rangle$ and $\langle \Delta_R, k_R\rangle$ by using (\ref{A6}) to write
\be
    \langle k_R^N k_L^n \Delta_R^M \Delta_L^m, k_R^W\rangle = \langle k_R^N \Delta_R^M , k_R^{N+M-m}\Delta_R^m \rangle = \langle k_R^N \Delta_R^{M-m} \Delta_R^m , k_R^{N+M-m}\Delta_R^m \rangle\,.
\ee
which can be further simplified by using (\ref{A5}) 
\begin{align}
    \langle k_R^N \Delta_R^{M-m} \Delta_R^m , k_R^{N+M-m}\Delta_R^m \rangle = \langle \Delta_R,\Delta_R\rangle^m \langle k_R,k_R \rangle^N \langle \Delta_R^{M-m},k_R^{M-m}\rangle\,.
\end{align}
This expression may appear strange for the case $M-m <0$ but it can be treated straightforwardly using (\ref{A6}) with the relation $\Delta_L = \delp^2/\Delta_R$. Eventually, we obtain
\begin{align}
    \langle k_R^N \Delta_R^{M-m} \Delta_R^m , k_R^{N+M-m}\Delta_R^m \rangle = \langle \Delta_R,\Delta_R\rangle^M \langle k_R,k_R \rangle^{N+M-m} \langle k_R^{m-M},\Delta_R^{m-M}\rangle\,.
\end{align}
An analogous procedure applies to $\langle k_R^N k_L^n \Delta_R^M \Delta_L^m, \Delta_R^W\rangle$. To summarize, we have
\begin{align}
        \langle k_R^N k_L^n \Delta_R^M \Delta_L^m, k_R^W\rangle =
    \begin{cases}
    \langle \Delta_R,\Delta_R\rangle^m \langle k_R,k_R \rangle^N \langle \Delta_R^{M-m},k_R^{M-m}\rangle, & \text{for } M-m \ge 0\,, \\
    \langle \Delta_R,\Delta_R\rangle^M \langle k_R,k_R \rangle^{N+M-m} \langle k_R^{m-M},\Delta_R^{m-M}\rangle,  & \text{for } M-m < 0\,,
    \end{cases}
\end{align}
and
\begin{align}
    \langle k_R^N k_L^n \Delta_R^M \Delta_L^m, \Delta_R^W\rangle =
    \begin{cases}
    \langle \Delta_R,\Delta_R\rangle^M \langle k_R,k_R \rangle^n \langle k_R^{N-n},\Delta_R^{N-n}\rangle, & \text{for } N-n \ge 0\,, \\
    \langle \Delta_R,\Delta_R\rangle^{M+N-n} \langle k_R,k_R \rangle^{N} \langle \Delta_R^{n-N},k_R^{n-N}\rangle,  & \text{for } N-n < 0\,.
    \end{cases}
\end{align}

%\begin{align}
%\langle k_R^N k_L^n \Delta_R^M \Delta_L^m, k_R^W\rangle&= \langle k_R^N \Delta_R^M , k_R^{N+M-m}\Delta_R^m \rangle \nonumber\\
%&= \langle \Delta_R, \Delta_R \rangle^{\min(M,m)} \langle k_R, k_R \rangle^{\min(N,N+M-m)} \nonumber\\
%    & \ \ \ \times\langle k_R^{\max(0,m-M)} \Delta_R^{\max(0,M-m)} ,  k_R^{\max(0,M-m)} \Delta_R^{\max(0,m-M)} \rangle , \\
 %\langle k_R^N k_L^n \Delta_R^M \Delta_L^m, \Delta_R^W\rangle &= \langle k_R^N \Delta_R^M , k_R^n\Delta_R^{N-n+M} \rangle \nonumber\\
 %&=\langle k_R, k_R\rangle^{\min(N,n)} \langle \Delta_R, \Delta_R\rangle^{\min(M,M+N-n)} \nonumber\\
  %  & \ \ \ \times\langle k_R^{\max(0,N-n)} \Delta_R^{\max(0,n-N)} ,  k_R^{\max(0,n-N)} \Delta_R^{\max(0,N-n)} \rangle .
%\end{align}
%Thus, all coefficients can be expressed in terms of $\langle k_R, k_R\rangle,\langle \Delta_R, \Delta_R\rangle, \langle \Delta_R, k_R\rangle$.

We now illustrate the derived procedure through several examples:
\begin{enumerate}
    \item[1)] $N=2,n=0,M=0,m=1, W=1$
\begin{align}
    k_R^2 \Delta_L = & \frac{ \langle \Delta_R, \Delta_R \rangle \langle k_R, k_R \rangle \langle k_R , \Delta_R \rangle - \langle k_R^2 , \Delta_R^2 \rangle \langle k_R , \Delta_R \rangle }{\langle \Delta_R, \Delta_R \rangle \langle k_R, k_R \rangle - \langle k_R , \Delta_R \rangle^2} k_R \nonumber\\
    &+\frac{- \langle k_R, k_R \rangle \langle k_R , \Delta_R \rangle\langle k_R , \Delta_R \rangle +\langle k_R, k_R \rangle  \langle k_R^2 , \Delta_R^2 \rangle  }{\langle \Delta_R, \Delta_R \rangle \langle k_R, k_R \rangle - \langle k_R , \Delta_R \rangle^2} \Delta_R \nonumber\\
    =& 2 \langle k_R, \Delta_R \rangle k_R - \langle k_R, k_R \rangle \Delta_R\,.
\end{align}
    \item[2)] $N=1,n=0,M=1,m=0, W=2$
\begin{align}
    k_R \Delta_R =& \frac{\langle \Delta_R, \Delta_R \rangle^2 \langle k_R, k_R \rangle \langle \Delta_R, k_R \rangle - \langle \Delta_R, \Delta_R \rangle \langle \Delta_R, k_R \rangle\langle \Delta_R^2, k_R^2 \rangle}{\langle \Delta_R, \Delta_R \rangle^2 \langle k_R, k_R \rangle^2 - \langle \Delta_R^2, k_R^2 \rangle^2} k_R^2 \nonumber\\
    &+\frac{- \langle k_R, k_R \rangle \langle \Delta_R, k_R \rangle\langle \Delta_R^2, k_R^2 \rangle + \langle k_R, k_R \rangle^2 \langle \Delta_R, \Delta_R \rangle \langle \Delta_R, k_R \rangle}{\langle \Delta_R, \Delta_R \rangle^2 \langle k_R, k_R \rangle^2 - \langle \Delta_R^2, k_R^2 \rangle^2} \Delta_R^2 \nonumber\\
    =&\frac{\langle \Delta_R,\Delta_R \rangle}{2 \langle k_R, \Delta_R \rangle} k_R^2 + \frac{\langle k_R,k_R \rangle}{2 \langle k_R, \Delta_R \rangle} \Delta_R^2\,.
\end{align}
    \item[3)]  $N=2,n=0,M=1,m=0, W=3$
\begin{align}
    k_R^2 \Delta_R =& \frac{\langle \Delta_R, \Delta_R \rangle^3 \langle k_R, k_R \rangle^2 \langle \Delta_R, k_R \rangle - \langle \Delta_R, \Delta_R \rangle \langle \Delta_R^2, k_R^2 \rangle\langle \Delta_R^3, k_R^3 \rangle}{\langle \Delta_R, \Delta_R \rangle^3 \langle k_R, k_R \rangle^3 - \langle \Delta_R^3, k_R^3 \rangle^2} k_R^2 \nonumber\\
    &+\frac{- \langle k_R, k_R \rangle^2 \langle \Delta_R, k_R \rangle\langle \Delta_R^3, k_R^3 \rangle + \langle k_R, k_R \rangle^3 \langle \Delta_R, \Delta_R \rangle \langle \Delta_R^2, k_R^2 \rangle}{\langle \Delta_R, \Delta_R \rangle^3 \langle k_R, k_R \rangle^3 - \langle \Delta_R^3, k_R^3 \rangle^2} \Delta_R^2 \nonumber\\
    =&\frac{2\langle \Delta_R,\Delta_R \rangle \langle k_R,\Delta_R \rangle}{4 \langle k_R, \Delta_R \rangle^2 - \langle \Delta_R,\Delta_R \rangle \langle k_R,k_R \rangle} k_R^3 + \frac{\langle k_R,k_R \rangle^2}{4 \langle k_R, \Delta_R \rangle^2 - \langle \Delta_R,\Delta_R \rangle \langle k_R,k_R \rangle} \Delta_R^3\,.
\end{align}
\end{enumerate}

\section{Intermediate computation steps for sea-quark GTMDs}
\label{sec:steps}

{Here we present the derivations of \eqref{eq:isotrop}, \eqref{eq:ellipt}, and \eqref{eq:Lflip2}. The general idea is based on the fact that $A$ from \eqref{eq:Aconst}, together with $B_1$ and $B_2$ from \eqref{eq:B12}, are analytic functions of $\cos(\phi_{gk})$. As such, they can be expanded in Fourier series for $\cos(\phi_{gk})$, leading to the decomposition
\be
A=\sum_{n=0}^\infty A_n\cos(n\phi_{gk})\,,
\label{eq:decoma}
\ee
with the coefficients
\be
A_0 = \frac{1}{2\pi} \int_0^{2\pi} \rmd \phi_{gk}\, A\,,\qquad A_n = \frac{1}{\pi} \int_0^{2\pi} \rmd \phi_{gk}\, A \cos(n\phi_{gk})\,,
\label{eq:coef}
\ee
and analogous expressions for $B_1$ and $B_2$. By inserting \eqref{eq:decoma} into the LHS of \eqref{eq:isotrop}, we obtain
\be
\int\rmd^2 \kgp A \frac{k_{gL}}{M} \frac{\kgp\cdot\delp}{M^2} \calP^\perp_{1T,0}
=\sum_{n=0}^\infty\int_0^\infty\frac{\rmd\kgp^2}{2} \frac{\kgp^2|\delp|}{M^3}\calP^\perp_{1T,0}A_n
\int_0^{2\pi}\rmd\phi_g\rme^{-\rmi\phi_g}\cos(n\phi_{gk})\cos(\phi_{g\Delta})\,.
\label{eq:I1}
\ee
We now evaluate this angular integral and then, with the help of \eqref{eq:coef}, re-express all coefficients $A_n$ that survive. We find
\be
\begin{split}
    \sum_{n=0}^\infty A_n&\int_0^{2\pi}\rmd\phi_g\rme^{-\rmi\phi_g}\cos(n\phi_{gk})\cos(\phi_{g\Delta})
    =\sum_{n=0}^\infty A_n\pi\left[\rme^{-\rmi\phi_\Delta}\left(\delta_{n0}-\frac{\delta_{n2}}{2}\right)
    +\rme^{-\rmi\phi_k}\cos(\phi_{\Delta k})\delta_{n2}\right]\\
    &=\pi\left[\rme^{-\rmi\phi_\Delta}\left(A_0-\frac{A_2}{2}\right)
    +\rme^{-\rmi\phi_k}\cos(\phi_{\Delta k})A_2\right]\\
    &=\frac{1}{2}\left[\rme^{-\rmi\phi_\Delta}\int_0^{2\pi}\rmd\phi_{gk}A\left(1-\cos(2\phi_{gk})\right)
    +\rme^{-\rmi\phi_k}\cos(\phi_{\Delta k})\int_0^{2\pi}\rmd\phi_{gk}A\cos(2\phi_{gk})\right]\\
    &=\frac{\Delta_L}{|\delp|}\int_0^{2\pi}\rmd\phi_{gk}A\left[1-\frac{(\kp\cdot\kgp)^2}{\kp^2\kgp^2}\right]
    +\frac{k_L}{|\kp|}\frac{\delp\cdot\kp}{|\delp||\kp|}
    \int_0^{2\pi}\rmd\phi_{gk}A\left[\frac{2(\kp\cdot\kgp)^2}{\kp^2\kgp^2}-1\right]\,.
\end{split}
\label{eq:ang}
\ee
By inserting the last line of \eqref{eq:ang} into the RHS of \eqref{eq:I1}, we immediately obtain \eqref{eq:isotrop}.

Although the derivation of \eqref{eq:ellipt} is a little bit more involved, the general idea is the same. We first insert \eqref{eq:decoma} into the LHS of \eqref{eq:ellipt} to obtain
\be
\begin{split}
\int\rmd^2\kgp A \frac{k_{gL}}{M} \frac{\kgp\cdot\delp}{M^2} &\left[\frac{2(\kgp\cdot\delp)^2}{\kgp^2 \delp^2} - 1\right] 2\calP^\perp_{1T,\epsilon}
= \sum_{n = 0}^\infty \int_0^\infty \frac{\rmd \kgp^2}{2} 2\calP_{1T,\epsilon}^\perp \frac{\kgp^2 |\delp|}{M^3}\\
& \times A_n \int_{0}^{2\pi} \rmd \phi_g \rme^{-\rmi \phi_g} \cos(\phi_{g\Delta}) \cos(2\phi_{g\Delta}) \cos(n \phi_{gk})\,.
\end{split}
\label{eq:initel1}
\ee
For the angular integral we find
\be
\begin{split}
    \sum_{n=0}^\infty& A_n\int_{0}^{2\pi} \rmd \phi_g \rme^{-\rmi \phi_g} \cos(\phi_{g\Delta}) \cos(2\phi_{g\Delta}) \cos(n \phi_{gk})\\
    &=\rme^{-\rmi \phi_k}\frac{\pi}{2}\left[\cos(\phi_{k\Delta}) + \cos(3\phi_{k\Delta})\right]A_4
    +\rme^{-\rmi \phi_\Delta} \frac{\pi}{2}\left[A_0 + \cos(2\phi_{k\Delta})A_2 - \frac{1}{2}(1 + 2 \cos(2\phi_{k\Delta}))A_4\right]\,.
\end{split}
\ee
As in \eqref{eq:ang}, we wish to express everything in terms of powers of $\cos(\phi_{\Delta k})$ and $\cos(\phi_{gk})$. We need the following Chebyshev polynomial identities:
\be
\begin{split}
    \cos(2x)&=2\cos^2(x)-1\\
    \cos(3x)&=4\cos^3(x)-3\cos(x)\\
    \cos(4x)&=8\cos^4(x)-8\cos^2(x)+1\,.
\end{split}
\ee
With this we have
\be
\begin{split}
    \sum_{n=0}^\infty& A_n\int_{0}^{2\pi} \rmd \phi_g \rme^{-\rmi \phi_g} \cos(\phi_{g\Delta}) \cos(2\phi_{g\Delta}) \cos(n \phi_{gk})\\
    &=\frac{k_L}{|\delp|}\frac{\kp\cdot\delp}{\kp^2}\left[\frac{2(\kp\cdot\delp)^2}{\kp^2\delp^2}-1\right]
    \int_0^{2\pi}\rmd\phi_{gk}A\left[\frac{8(\kp\cdot\kgp)^4}{\kp^4\kgp^4}-\frac{8(\kp\cdot\kgp)^2}{\kp^2\kgp^2}+1\right]\\
    &+\frac{\Delta_L}{|\delp|}\left[\frac{2(\kp\cdot\delp)^2}{\kp^2\delp^2}-1\right]
    \int_0^{2\pi}\rmd\phi_{gk}A\left[-\frac{4(\kp\cdot\kgp)^4}{\kp^4\kgp^4}+\frac{5(\kp\cdot\kgp)^2}{\kp^2\kgp^2}-1\right]\\
    &+\frac{\Delta_L}{|\delp|}\int_0^{2\pi}\rmd\phi_{gk}A\frac{2(\kp\cdot\kgp)^2}{\kp^2\kgp^2}\left[1-\frac{(\kp\cdot\kgp)^2}{\kp^2\kgp^2}\right]\,.
\end{split}
\label{eq:bigint}
\ee
Inserting \eqref{eq:bigint} into \eqref{eq:initel1} we obtain \eqref{eq:ellipt}.

As for \eqref{eq:Lflip2}, we start from Eq.~\eqref{eq:Lflip} and proceed with the same strategy. Our convention for the cross product (see below Eq.~\eqref{eq:GTMD1}) implies that
$
\boldsymbol{a}\times \boldsymbol{b} = |\boldsymbol{a}||\boldsymbol{b}|\sin(\phi_{ba}) \, .
$
We first calculate the part of \eqref{eq:Lflip} containing $B_1$. We have:
\be
\begin{split}
    2N_c&\int\rmd^2\kgp 2(\kp\cdot\delp)\frac{\qp\times\kp}{\kp^2}B_1\frac{k_{gL}}{M}\calO_{1T,0}^\perp\\
    &=4N_c\frac{\kp\cdot\delp}{\kp^2}\int_0^\infty\frac{\rmd \kgp^2}{2}\frac{|\kp|}{M}\kgp^2\calO_{1T,0}^\perp\sum_{n=0}^\infty B_{1,n}\int_0^{2\pi}\rmd\phi_g\sin(\phi_{gk})\cos\left(n\phi_{gk}\right)\rme^{-\rmi\phi_g}\\
    &=2\rmi N_c\frac{\kp\cdot\delp}{\kp^2}\frac{k_L}{M}\int\rmd^2\kgp\kgp^2\calO_{1T,0}^\perp\left[\frac{(\kp\cdot\kgp)^2}{\kp^2\kgp^2}-1\right]2B_1\,.
\end{split}
\label{eq:b1last}
\ee
Since the part of \eqref{eq:Lflip} proportional to $B_2$ contains a cross product $\qp\times\delp=\kp\times\delp-\kgp\times\delp$, this term is naturally calculated by separating these two contributions:
\be
\begin{split}
    2N_c&\int\rmd^2\kgp(\qp\times\delp)B_2\frac{k_{gL}}{M}O_{1T,0}^\perp\\
    &=2N_c(\kp\times\delp)\int_0^{\infty}\frac{\rmd \kgp^2}{2}\frac{|\kgp|}{M}O_{1T,0}^\perp\sum_{n=0}^{\infty}B_{2,n}\int_0^{2\pi}\rmd\phi_g\rme^{-\rmi\phi_g}\cos(n\phi_{gk})\\
    &-2N_c\frac{|\delp|}{M}\int_0^{\infty}\frac{\rmd \kgp^2}{2}\kgp^2 O_{1T,0}^\perp\sum_{n=0}^{\infty}B_{2,n}\int_0^{2\pi}\rmd\phi_g\rme^{-\rmi\phi_g}\sin(\phi_{\Delta g})\cos(n\phi_{gk})\\
    &=2N_c(\kp\times\delp)\frac{k_L}{M}\int\rmd^2\kgp O_{1T,0}^\perp\frac{\kp\cdot\kgp}{\kp^2}B_2-2\rmi N_c\frac{\Delta_L}{M}\int\rmd^2\kgp O_{1T,0}^\perp \frac{(\kp\cdot\kgp)^2}{\kp^2} B_2\\
    &+2\rmi N_c\frac{k_L}{M}\int\rmd^2\kgp\kgp^2O_{1T,0}^\perp \frac{\kp\cdot\delp}{\kp^2}\left[\frac{2(\kp\cdot\kgp)^2}{\kp^2\kgp^2}-1\right]B_2\,.
\end{split}
\ee
To eliminate cross product we use \eqref{eq:epsid} and get
\be
\begin{split}
    2N_c\int\rmd^2\kgp(\qp\times\delp)B_2\frac{k_{gL}}{M}O_{1T,0}^\perp&=2\rmi N_c\frac{\Delta_L}{M}\int\rmd^2\kgp \kgp^2 O_{1T,0}^\perp \left[\frac{\kp\cdot\kgp}{\kgp^2}-\frac{(\kp\cdot\kgp)^2}{\kp^2\kgp^2}\right]B_2\\
    &+2\rmi N_c\frac{\delp\cdot\kp}{\kp^2}\frac{k_L}{M}\int\rmd^2\kgp \kgp^2 O_{1T,0}^\perp \left[\frac{2(\kp\cdot\kgp)^2}{\kp^2\kgp^2}-\frac{\kp\cdot\kgp}{\kgp^2}-1\right]B_2\,.
\end{split}
\label{eq:b2last}
\ee
Combining \eqref{eq:b1last} and \eqref{eq:b2last} we obtain \eqref{eq:Lflip2}.
}

\typeout{}
\bibliography{references}

\end{document}